\documentclass[preprint2]{proto}
\usepackage{times}

\newcommand{\gsim}{\mathrel{\hbox{\rlap{\lower.55ex \hbox {$\sim$}}
                   \kern-.3em \raise.4ex \hbox{$>$}}}}
\newcommand{\lsim}{\mathrel{\hbox{\rlap{\lower.55ex \hbox {$\sim$}}
                   \kern-.3em \raise.4ex \hbox{$<$}}}}
\newcommand{\msun}{{M_\odot}}
\newcommand{\rsun}{{R_\odot}}
\newcommand{\lsun}{{L_\odot}}

\usepackage{color}
\newcommand{\red}[1]{{#1}}

\begin{document}

\title{\textbf{\LARGE Star Cluster Formation and Feedback}}

\author {\textbf{\large Mark R.~Krumholz}}
\affil{\small\em University of California, Santa Cruz}

\author {\textbf{\large Matthew R.~Bate}}
\affil{\small\em University of Exeter}

\author {\textbf{\large H\'{e}ctor G.~Arce}}
\affil{\small\em Yale University}

\author {\textbf{\large James E.~Dale}}
\affil{\small\em Excellence Cluster `Universe', Ludwig-Maximillians-University}

\author {\textbf{\large Robert Gutermuth}}
\affil{\small\em University of Massachusetts, Amherst}

\author {\textbf{\large Richard I.~Klein}}
\affil{\small\em University of California, Berkeley and Lawrence Livermore National Laboratory}

\author {\textbf{\large Zhi-Yun Li}}
\affil{\small\em University of Virginia}

\author {\textbf{\large Fumitaka Nakamura}}
\affil{\small\em National Astronomical Observatory of Japan}

\author {\textbf{\large Qizhou Zhang}}
\affil{\small\em Harvard-Smithsonian Center for Astrophysics}

\begin{abstract}
\baselineskip = 11pt
\leftskip = 0.65in 
\rightskip = 0.65in
\parindent=1pc
{\small 
Stars do not generally form in isolation. Instead, they form in clusters, and in these clustered environments newborn stars can have profound effects on one another and on their parent gas clouds. Feedback from clustered stars is almost certainly responsible for a number of otherwise puzzling facts about star formation: that it is an inefficient process that proceeds slowly when averaged over galactic scales; that most stars disperse from their birth sites and dissolve into the galactic field over timescales $\ll 1$ Gyr; and that newborn stars follow an initial mass function (IMF) with a distinct peak in the range $0.1 - 1$ $\msun$, rather than an IMF dominated by brown dwarfs. In this review we summarize current observational constraints and theoretical models for the complex interplay between clustered star formation and feedback.
 \\~\\~\\~}

\end{abstract}

\section{\textbf{INTRODUCTION}}

\noindent
\subsection{\textbf{Why is Feedback Essential?}}
\label{ssec:why}
\bigskip

\begin{figure*}[t!]
\epsscale{1.4}
\plotone{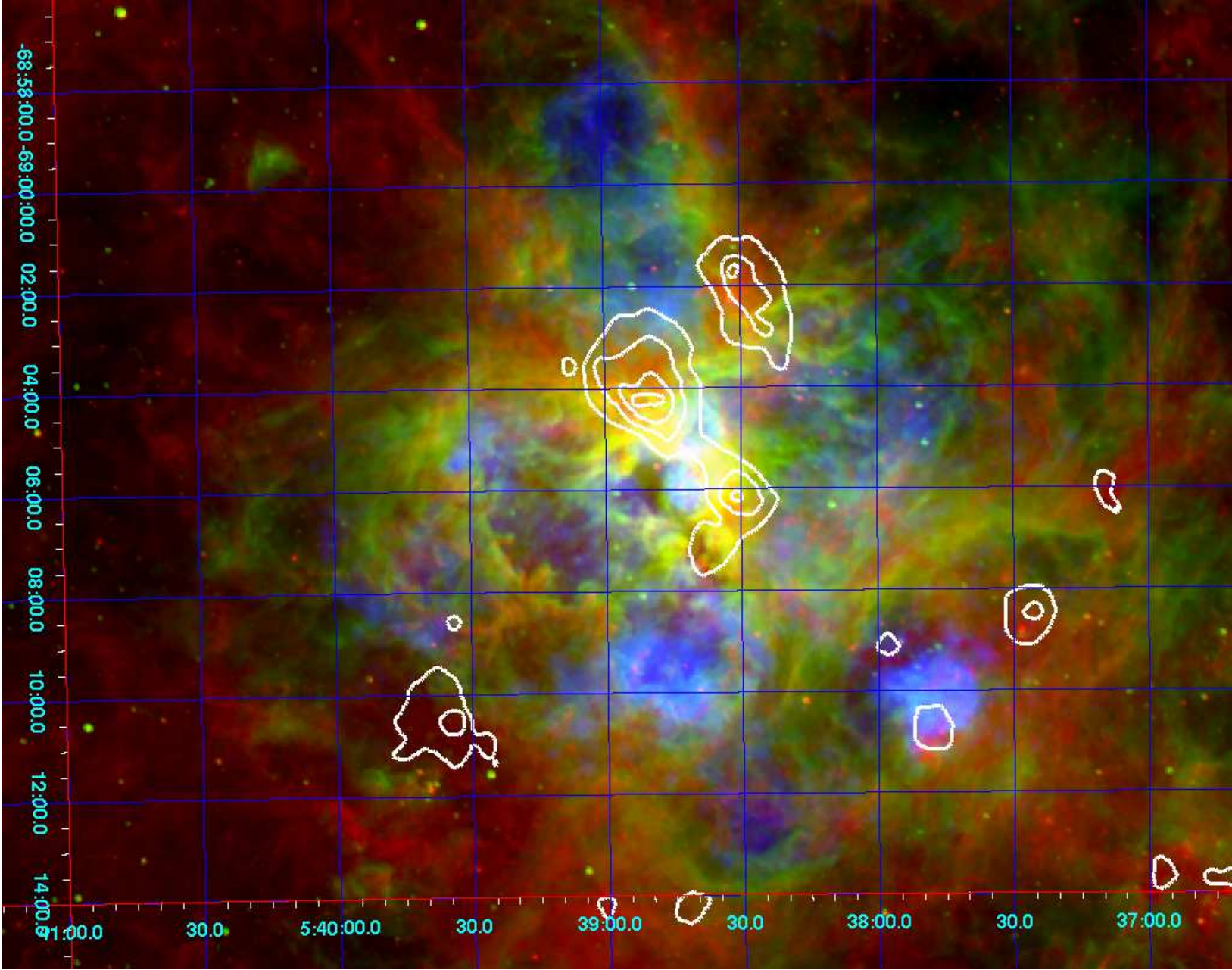}
\epsscale{1.0}
\caption{\small
\label{fig:30dor}
Three-color image of 30 Doradus: 8 $\mu$m (red), H$\alpha$ (green), and $0.5-8$ keV X-rays (blue). White contours show $^{12}$CO(1-0) emission. Figure taken from \citet{lopez11a}.
}
\end{figure*}

Newborn stars have profound effects on their birth environments, and any complete theory for star formation must include them. Perhaps the best argument for this statement is an image such as Figure \ref{fig:30dor}, which shows 30 Doradus, the largest H~\textsc{ii} region in the Local Group, powered by the cluster NGC 2070 and its 2400 OB stars \citep{parker93a}. The figure illustrates several of the routes by which young stars can influence their surroundings. The red color shows 8 $\mu$m emission, marking where gas has been warmed by \red{far ultraviolet} radiation \red{from young stars}. The green color traces H$\alpha$, indicating where ionizing radiation has converted the interstellar medium to a warm ionized phase. Finally, blue shows X-ray emission from a $\sim 10^7$ K phase created by shocks in the fast winds launched by the O stars. \red{The entire region is expanding at $\sim 25$ km s$^{-1}$ \citep{chu94a}}. Any theory for how NGC 2070 arrived at its present state must address the role played by these processes, and several others \red{such as protostellar outflow feedback and radiative heating by infrared light. These} are \red{not} visually apparent in Figure \ref{fig:30dor}, but \red{can be seen clearly in other regions, and} are perhaps equally important.

In addition to the visual impression provided by Figure \ref{fig:30dor} and similar observations, there are a number of more subtle but equally compelling arguments for the importance of stellar feedback for cluster formation. The first, which dates back to the seminal work of \citet{zuckerman74a}, is based on comparing the molecular mass $M_{\rm gas}$ to the star formation rate $\dot{M}_*$ \red{-- either in an entire galaxy, or in a smaller region defined by a specified volume or column density threshold --} to deduce a characteristic depletion time $t_{\rm dep} = M_{\rm gas}/\dot{M}_*$. Note that this should not be confused with the time scale over which star formation in a particular cloud takes place, as diagnosed for example by the stellar age spread (see the chapter by \textit{Soderblom et al.}~in this volume). The two are identical only if star formation proceeds until all the gas is converted to stars, and in general the depletion time is $1-2$ orders of magnitude larger.

While \citeauthor{zuckerman74a} applied this technique to the low-density molecular gas traced by low-$J$ CO emission, and numerous modern studies have done the same (see the chapters by \textit{Dobbs et al.}~and \textit{Padoan et al.}~in this volume), it has also become possible in the last ten years to perform the same analysis for tracers of the denser regions from which clusters presumably form. \red{Techniques for studying such regions include the low-$J$ lines of heavy rotor molecules such as HCN, HCO$^+$, and CS (which have critical densities $\gtrsim 10^4$ cm$^{-3}$), thermal emission from cold dust at sub-mm wavelengths, and dust extinction at near-infrared wavelengths.} The consensus from such studies is that the depletion time is always $\sim 1- 3$ orders of magnitude longer than the free-fall time $t_{\rm ff} \sim 1/\sqrt{G\rho}$, where $\rho$ is the characteristic density selected by the tracer \citep{krumholz07b, evans09a, juneau09a, krumholz12a, federrath13c}. In contrast, numerical simulations of star cluster formation that do not include any form of feedback generally produce $t_{\rm dep} \sim t_{\rm ff}$ \citep[e.g.,][]{klessen00a, klessen01a, bate03a, bonnell03a}.

Magnetic fields alone are unlikely to prevent this outcome. Observations suggest that the median cloud is magnetically supercritical by a factor of 2 \citep[and references therein]{crutcher12a}, and simulations indicate that a magnetic field of this strength only decreases the star formation rate by a factor of a few compared to the purely hydrodynamic case (\citealt{PriBat2009, padoan11a, federrath12a}; also see the chapter by \textit{Padoan et al.}~in this volume). Reduction of the star formation rate by feedback, perhaps in conjunction with magnetic fields, is a prime candidate to resolve this problem.

A second, closely related argument has to do with the fraction of stars found in bound clusters. Most regions of active star formation are much denser than the field \citep[e.g.,][]{gutermuth09a}. \red{We will refer to these regions as clusters, defined roughly as suggested by \citet{lada03a}: a collection of physically-related stars within which the stellar mass density is $\gg 1$ $M_\odot$ pc$^{-3}$ (compared to $\sim 0.01$ in the field near the Sun -- \citealt{holmberg00a}), and where the total number of stars is greater than several tens. Such regions are typically $\sim 1-10$ pc in size, and, at least when they are young, also contain gas with a mass density greatly exceeding that of the stars. We do not require, as do some authors interested primarily in N-body dynamics \citep[e.g.,][]{portegies-zwart10a}, that the stars in question be gravitationally bound, or old enough to be dynamically-relaxed; the former condition is often impossible to evaluate in clusters that are still embedded in their natal gas clouds, while the latter necessarily excludes the phase of formation in which we are most interested.}

\red{The argument for the importance of feedback can be made by observing that, while almost all regions of active star formation qualify as clusters by this definition,} by an age of $\sim 10-100$ Myr, only a few percent of stars remain part of clusters with stellar densities noticeably above that of the field \cite[e.g.,][]{silva-villa11a, fall12a}. In a cluster of $N$ stars with a crossing time $t_{\rm cross}$ (typically $\sim 0.1-1$ Myr in observed clusters), two-body evaporation does not become important until an age of $(10 N/\ln N) t_{\rm cross}$ \citep{binney87a}. This is $\sim 100-1000$ Myr even for a modest cluster of $N = 1000$. Thus \red{a gravitationally-bound cluster will not disperse on its own over the timescale demanded by observations.} However, simulations of star cluster formation that do not include feedback have a great deal of difficulty reproducing this outcome \citep{bate03a, bonnell03a}. Instead, they invariably produce bound clusters.

\red{One might think that this problem could be avoided by positing that most stars form in gravitationally-unbound gas clouds. However, such a model suffers from two major problems. First, as discussed in the chapter by \textit{Dobbs et al.}~in this volume, recent surveys of molecular clouds find that their typical virial ratios are $\alpha_G \approx 1$, whereas $\alpha_G>2$ is required to render a cloud unbound. Second, \citet{clark05b} find that even unbound clouds (they consider one with $\alpha_G=4$) leave most of their stars in bound clusters. A high virial ratio means that a cloud produces a number of smaller, mutually-unbound clusters rather than a single large one, but each sub-cluster is still internally bound and has $>1000$ stars. These would survive too long to be consistent with observations. Stellar feedback represents the most likely way out of this problem, as the dispersal of gas by feedback can reduce the star formation efficiency to the point where few stars remain members of bound clusters.}

A final argument for the importance of feedback comes from the problem of explaining the origin of the stellar initial mass function (IMF). As we discuss below, simulations that do not include radiative feedback tend to have problems reproducing the observed IMF, while those including it do far better (also see the chapter by \textit{Offner et al.}). In the following sections, we discuss the role of feedback in solving each of these problems, and highlight both the successes and failures of current models for its operation.

\bigskip
\noindent
\subsection{\textbf{A Taxonomy of Feedback Mechanisms}}
\label{ssec:taxonomy}
\bigskip

Before discussing individual feedback mechanisms in detail, it is helpful to lay out some categories that can be used to understand them. Although many such taxonomies are possible, we choose to break feedback mechanisms down into three categories: momentum feedback, ``\red{explosive}" feedback, and thermal feedback.

Momentum feedback is, quite simply, the deposition of momentum into star-forming clouds so as to push on the gas, drive turbulent motions within it, and, if the feedback is strong enough, to unbind them entirely. The key feature of momentum feedback, which distinguishes it from \red{explosive} feedback, is the role of radiative energy loss. The dense, molecular material from which stars form, or even the less dense gas of the atomic ISM, is extremely efficient at radiative cooling. As a result, when stars inject energy into the ISM, it is often the case that the energy is then radiated away on a timescale that is short compared to the dynamical time of the surrounding cloud. In this case the amount of energy delivered to the cloud matters little, and the effectiveness of the feedback is instead determined by the amount of momentum that is injected, since this cannot be radiated away. As we discuss below, protostellar outflows and (probably) radiation pressure are forms of momentum feedback.

In contrast, \red{explosive} feedback occurs when stars heat gas so rapidly, and to such a high temperature, that it is no longer able to cool on a cloud dynamical timescale. In this case at least some of the energy added to the cloud is not lost to radiation, and feedback is accomplished when the hot, overpressured gas expands \red{explosively} and does work on the surrounding cold molecular material. To understand the distinction between \red{explosive} and momentum-driven feedback, consider a point source injecting a wind of material into a uniform, cold medium, and sweeping up an expanding shell of material of mass $M_{\rm sh}$ and radius $r_{\rm sh}$. If the wind is launched with mass flux $\dot{M}_w$ at velocity $V_w$, and in the process of sweeping up the shell there are no radiative losses (the extreme limit of the \red{explosive} case), then \red{after a time $t$ the kinetic energy of the shell is} $M_{\rm sh} \dot{r}_{\rm sh}^2 \sim \dot{M}_w V_w^2 \red{t}$. On the other hand, in the case of momentum feedback where energy losses are maximal, \red{the momentum of the shell is $M_{\rm sh} \dot{r}_{\rm sh} \sim \dot{M}_w V_w t$, and its kinetic energy is $M_{\rm sh} \dot{r}_{\rm sh}^2 \sim \dot{M}_w V_w \dot{r}_{\rm sh} t$}. Thus without radiative losses, \red{the kinetic energy of the shell at equal times is larger by a factor of $\sim V_w/\dot{r}_{\rm sh}$}. This is not a small number: in the example of 30 Doradus (Figure \ref{fig:30dor}), the measured velocity of the shell is $\sim 25$ km s$^{-1}$ \citep{chu94a}, while typical launching velocities for O star winds are $>1000$ km s$^{-1}$. Thus, when it operates, \red{explosive} feedback can be very effective. Winds from hot main sequence stars, photoionizing radiation, and supernovae are all forms of \red{explosive} feedback.

Our final category is thermal feedback, which describes feedback mechanisms that do not necessarily cause the gas to undergo large-scale flows, but do alter its temperature. This is significant because the temperature structure of interstellar gas is strongly linked to how it fragments, and thus to the production of the IMF. Non-ionizing radiation is the main form of thermal feedback.

\bigskip
\section{\textbf{MOMENTUM FEEDBACK}}

\noindent
\subsection{\textbf{Protostellar Outflows}}
\label{ssec:outflows}
\bigskip

\subsubsection{Theory}
\label{sssec:outflowtheory}

Protostellar outflows are observed to be an integral part of star formation. Outflows eject a significant amount of mass from the regions around newborn stars, thereby helping to set the most important quantity for individual stars, their mass. Collectively, the outflows inject energy and momentum into their surroundings, modifying the environment in which the stars form \citep{norman80a, mckee89a, shu99a}. Here we focus on this interaction, leaving the question of wind launching mechanisms to the chapter by \textit{Frank et al.} A key issue for outflows, as for all feedback mechanisms, is their momentum budget per unit  mass of stars formed (a quantity with units of velocity), which we denote $V_{\rm out}$. \red{Note that $V_{\rm out}$ is \textit{not} the velocity with which an individual outflow is launched, it is the total momentum carried by the outflows divided by the total mass of stars formed. It is therefore smaller than the velocity of an individual outflow by a factor equal to the ratio of the mass injected into the outflow divided by the stellar mass formed.} A number of authors have attempted to estimate $V_{\rm out}$ from both theoretical models of outflow launching and from observed scaling relationships between outflow momenta and stellar properties. \citet{matzner00a} and \citet{matzner07a} estimate $V_{\rm out} = 20-40$ km s$^{-1}$. The bulk of the momentum is produced by low-mass stars rather than massive ones, because outflow launch speeds scale roughly with the escape speeds from stellar surfaces, and the escape speeds from high mass stars are not larger than those from low mass stars by enough to compensate for the vastly greater mass contained in low mass stars.

Protostellar outflow feedback is expected to be especially important wherever a large number of stars form close together in both space and time. The paradigmatic object for this type of feedback is the low-mass protocluster NGC 1333, where molecular line and infrared observations reveal numerous outflows packed closely together \citep{knee00a, walawender05a, Walawender08a, curtis10a, plunkett13a}. The significance of outflow feedback in cluster formation can be illustrated using a simple estimate. Let the mass of the stars in a cluster be $M_*$, so the total momentum injected into the cluster-forming clump is $M_* V_{\rm out}$. This momentum is in principle enough to move all of the clump material (of total mass $M_c$) by a speed of 
\begin{equation}
v \sim \mbox{SFE}\times V_{\rm out} \sim 5
\mbox{ km/s} \left({\mbox{SFE}\over 0.2}\right) \left({V_{\rm out}\over 25 \mbox{ km s}^{-1}}\right)
\end{equation}
where $\mbox{SFE}=M_*/M_c$ is the star formation efficiency of the clump. For typical parameters, this speed is significantly higher than
the velocity dispersion of low-mass protoclusters such as NGC 1333. If all of the momenta from the outflows were to be
injected simultaneously, they would unbind the clump completely. If they are injected gradually, they may maintain the turbulence in the clump against dissipation and keep the stars forming at a relatively low rate over several free-fall times. This slow star formation over an extended period is consistent with the simultaneous presence of objects in all evolutionary stages, from prestellar cores to evolved Class III objects that have lost most of their disks. The latter objects should be at least a few million years old, several times the typical free fall time of the dense clumps that form NGC 1333-like clusters \red{\citep{evans09a}}. 

Numerical simulations have demonstrated that outflows can indeed drive turbulence in a cluster-forming clump and maintain star formation well beyond one free-fall time. \citet{li06a} simulated magnetized cluster formation with outflow feedback assuming that stellar outflows are launched isotropically,
and showed that the outflows drive turbulence that keeps the cluster-forming clump in quasi-equilibrium. The same conclusion was reached independently by \citet{matzner07a}, who studied outflow-driven turbulence analytically. \citet{nakamura07a} showed that collimated outflows are even more efficient in driving turbulence than spherical ones, because they reach large distances and larger-scale turbulence tends to decay more slowly. \citet{banerjee07a} questioned this conclusion, performing simulations showing that fast-moving jets do not  excite significant supersonic motions in a smooth ambient medium. However, \citet{cunningham09a}  showed that jets running into a turbulent ambient medium are more efficient in driving turbulent motions. Indeed, \citet{carroll09a}  were able to demonstrate explicitly that fully developed turbulence can be driven and maintained by a collection of collimated outflows, even in the absence of any magnetic field. Magnetic fields tend to couple different parts of the clump material together, which enables the outflows to deposit their energy and momentum in the ambient medium more efficiently \citep{nakamura07a, wang10a}.

\begin{figure}[ht]
\epsscale{1.0}
\plotone{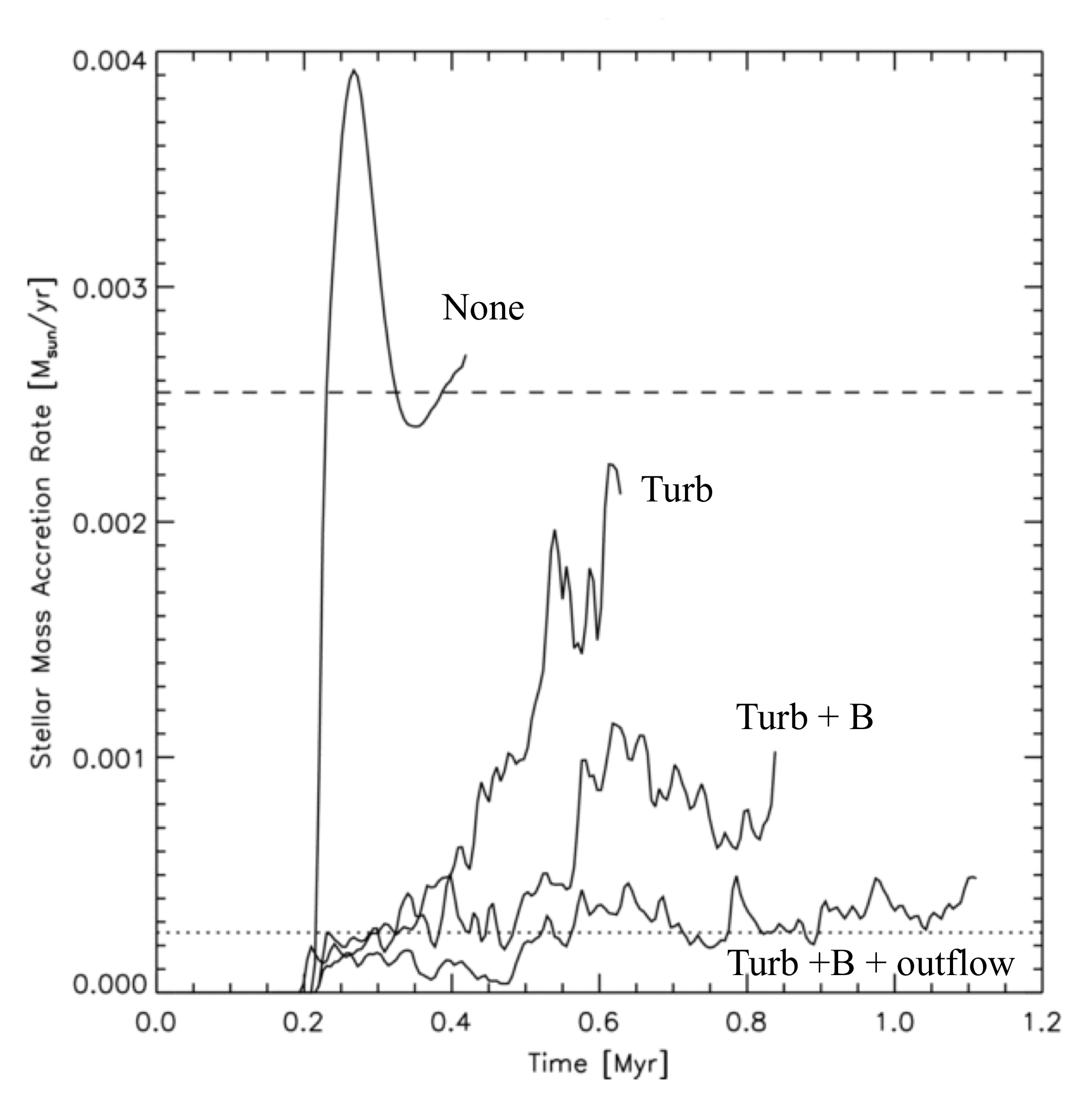}
\epsscale{1.0}
\caption{\small
\label{fig:wang}
The rates of star formation as a function of time
for four simulations of cluster formation \citep[adopted from][]{wang10a}. 
The curves from top left to bottom right are for models that 
include, respectively, neither turbulence nor magnetic field nor
outflow feedback (line labelled \textit{None}),
turbulence only (\textit{Turb}), turbulence and magnetic field
(\textit{Turb+B}), and 
all three ingredients (\textit{Turb+B+outflow}, see text for discussion). 
\red{The dashed horizontal line indicates a star formation rate such that the depletion
time $t_{\rm dep}$ is equal to the free-fall time $t_{\rm ff}$, while the dotted line indicates a star formation rate 
corresponding to $t_{\rm dep} = 10 t_{\rm ff}$.}
}
\end{figure}

The effects of magnetic fields and outflow feedback on cluster formation are illustrated in Figure \ref{fig:wang}, which shows the rates of star
formation in a parsec-scale clump with mass of order $10^3$ $\msun$ for four simulations of increasing complexity.  In the simplest case of no turbulence, magnetic field or outflow feedback (the top left line in the figure), the clump collapses rapidly, forming stars at a rate approaching  the characteristic free-fall rate (the dashed line). This rate is reduced progressively with the inclusion of initial turbulence (second solid line from top), turbulence and magnetic field (third line), and turbulence, magnetic field, and outflow feedback (bottom line). In the case with all three ingredients, the star formation rate is kept at $\sim 10\%$ of the free-fall rate (the dotted line). One implication of these and other simulations is that the majority of the cluster members may be formed in a relatively leisurely manner in an outflow-driven, magnetically-mediated, turbulent state, rather than rapidly in a free-fall time. Simple analytic estimates by \citet{nakamura11c} suggest that outflows should be able to maintain such low star formation rates in most of the observed clumps in the Solar neighborhood.

Another implication is that outflows from low-mass stars can influence the formation of the massive stars that form in the same cluster. For example, the same simulations that show a reduction in the star formation rate due to outflow feedback also show that outflows prevent rapid mass infall towards the massive stars that tend to reside at the bottom of the gravitational potential of protoclusters. Outflows can also have important interactions with other forms of feedback from both low mass and high mass stars, a topic we defer to Section \ref{sec:interactions}.

\subsubsection{Observations}

Several recent studies have searched for observational signatures of the outflow feedback effects described in the previous section. Even though not all studies use the same procedure to estimate the molecular outflow energetics and different studies use different methods to assess the combined impact from all outflows on the cluster gas, there are some points of consensus. One is that, in the vast majority of regions, the combined action of all outflows seems to be sufficient to drive the observed level of turbulence. The simplest method used in observational studies to determine this has been to compare the total mechanical energy of molecular outflows (i.e., the total kinetic energy of the molecular gas that has been entrained by the protostellar wind) with the turbulent energy of the cluster gas. Studies show that for a number of clusters the current total molecular outflow energy is $\sim 30\%$ or more of the total turbulent energy \citep[e.g.][]{arce10a, curtis10a, graves10a, duarte-cabral12a}, yet for other regions the total outflow energy is just $1-20\%$ of the total turbulent energy \citep[e.g.][]{arce10a, narayanan12a}. \red{It is unclear to what extent this differences in outflow energy are correlated with other cloud properties. \citet{sun06a} study CO(3-2) emission from different regions of Perseus, and find that those with active star formation, such as NGC 1333, show steeper velocity power spectral indices than more quiescent regions. However, it is not clear if differences in the turbulent energy injection are the cause, and the observations themselves are at relatively low ($\sim 80\arcsec$) resolution. Further study of this question is needed.}

Although such simple comparisons are useful to gauge the relative importance of outflows, they do not necessarily indicate whether outflows can \textit{maintain} the observed turbulence. A better way to address this is by comparing the total outflow power ($L_{\rm flow}$) or mechanical luminosity (i.e., the rate at which the outflows inject energy into their surroundings through entrainment of molecular gas by the protostellar wind) with the turbulent energy dissipation rate ($L_{\rm turb}$). Although there are many uncertainties associated with the estimation of both $L_{\rm flow}$ and $L_{\rm turb}$ from observations \citep[e.g.][]{williams03a, arce10a}, it is clear that for most protostellar clusters observed thus far  $L_{\rm flow} \sim L_{\rm turb}$ \citep{williams03a, stanke07a, swift08a, maury09a, arce10a, nakamura11a, nakamura11b}. The usual interpretation  has been that outflows have sufficient power to sustain (or at least provide a major source of power for maintaining) the turbulence in the region.

The physical assumption behind these observational comparisons is that the gas that has been put in motion through the interaction of the protostellar wind and the ambient medium (that is, the gas that makes up the bipolar molecular outflow) will eventually slow down and feed the turbulent motions of the cloud through some (not well understood) mechanism. However, it is unclear how efficiently outflow motions convert into cloud turbulence, and observational studies typically do not address this issue. Among the few exceptions are the studies by \citet{swift08a} and \citet{duarte-cabral12a}, which use observations of $^{13}$CO and C$^{18}$O (which are much better at tracing cloud structure than the more commonly observed $^{12}$CO) to investigate how outflows create turbulence. These studies show direct evidence of outflow-induced turbulence, but both conclude that  only a fraction of the outflow mechanical luminosity is used to sustain the turbulence in the cloud while a significant amount is deposited outside the cloud. They also suggest that the typical outflow energy injection scale, the scale at which the outflow momentum is most efficiently injected, is around a few tenths of a parsec, which agrees with the theoretical and numerical prediction \citep{matzner07a,nakamura07a}. However, the  clouds in these studies are relatively small and host large outflows. Similar observations of larger and denser clouds are needed to further investigate if the ``outflow-to-turbulence'' efficiency depends on cloud environment.  

We note that even though most studies show that outflows have the potential to have significant impact on the cluster environment, recent cloud-wide surveys have shown that outflows lack the power needed to sustain the observed turbulence on the scale of a molecular cloud complex \citep{walawender05a, arce10a, narayanan12a} or a giant molecular cloud \citep{dent09a, ginsburg11a}, with sizes of more than 10 pc. 
Outflows in cloud complexes and GMCs are mostly clustered in regions  with sizes of 1 to 4 pc, and there are large extents inside clouds with few or no outflows. This implies that an additional energy source is responsible for turbulence on a global cloud scale. See the chapter by \textit{Dobbs et al.}~in this volume for more discussion of this topic.

Several investigators have also performed observations to study the role of outflows in gas dispersal
\red{\citep[e.g.,][]{knee00a, swift08a, arce10a, curtis10a, graves10a, narayanan12a, plunkett13a}}.  
As with the work on turbulent driving, different studies use different methods to attack this problem. One common practice has been to compare the total kinetic energy of all molecular outflows with the cluster-forming clump's gravitational binding energy, with the typical result being that total outflow kinetic energy is less than 20\% of the binding energy \citep[e.g.,][]{arce10a, narayanan12a}. 
This simple analysis seems to indicate that in most regions outflows do not have enough energy to significantly disrupt their host clouds. An equivalent way of phrasing this conclusion is that, if all of the detected (current) outflow momentum were used to accelerate gas to the region's escape velocity, at most $5-10\%$ of the clump's mass could potentially be dispersed \citep{arce10a}. This is in stark contrast with the theoretical work of \citet{matzner00a}, which suggests ejection fractions of $50-70\%$. One possible explanation for the discrepancy is that a significant fraction of the outflow energy is deposited outside the cloud and is not detected by the observations. Another possibility is that the momentum in currently-visible outflows might be only a fraction of the total momentum from all outflows throughout the life of the cloud. However, the difference between current and total momentum would need to be a factor of $\sim 10$ to reconcile observations with theory, and a more likely scenario is that outflows disperse some of the gas, while other mechanisms, such as stellar winds and UV radiation \citep[e.g.,][]{arce11a} remove the rest. Certainly, further observations are needed in order to better understand the role of outflows in cluster gas dispersal. 
 
Finally, we caution that most detailed observations of outflows have concentrated on relatively nearby ($d < 500$ pc) clusters, which, for the most part, are only forming low to intermediate-mass stars. These regions do not accurately represent the galactic cluster population, introducing a bias. For example, the nearby regions do not reach the stellar densities or total stellar masses found in more distant regions.
There have been a number of outflow observations of high mass star-forming regions 
\citep[e.g.,][]{beuther02a, gibb04a, wu05a, zhang05a, lopez-sepulcre09a, sanchez-monge13a}.
However, such regions are typically more than $1-2$ kpc away that only the largest (and usually most powerful) outflows in each region are resolved.
High angular resolution is essential to untangle the emission from the numerous outflows present in high-mass forming regions
\citep[e.g.,][]{beuther03a,leurini09a, varricatt10a}.
Consequently, a complete census of outflows (especially those from low mass stars) in these far away clusters has been difficult with current instruments, and more detailed observations of high-mass star-forming regions are needed in order to better understand the role of outflows in clusters. Such observations will be complicated by the fact that more massive regions will contain massive stars that produce other forms of feedback, as discussed below. This will require disentangling outflows from these other effects. The best approach may be to focus on regions that are dense and massive, but are also relatively young and thus have formed few or no massive stars yet. ALMA will certainly be instrumental in conducting these studies.

\bigskip
\subsection{\textbf{Radiation Pressure}}
\label{ssec:radpressure}
\bigskip

\subsubsection{Theory and Radiation Trapping}

In clusters containing massive stars, a second form of momentum feedback comes into play: radiation pressure. Radiation feedback in general consists of the transfer of both energy and momentum from the radiation field generated by stars to the surrounding gas, but in this section we shall focus on the transfer of momentum. Except for photons above $13.6$ eV, this transfer is mostly mediated by dust grains. In comparison to protostellar outflows, which deliver a momentum \red{per unit mass of stars formed} $V_{\rm out} \sim 20-40$ km s$^{-1}$ (over a time comparable to the accretion time, $\sim 0.1$ Myr), a zero-age stellar population drawn from a fully-sampled IMF produces a radiation field of $1140$ $\lsun$ per $\msun$ of stars \citep{murray10b}, which carries a momentum \red{per unit time per unit stellar mass} $\dot{V}_{\rm rad} \sim 24$ km s$^{-1}$ Myr$^{-1}$. Thus, over the time $t_{\rm form} \sim 1 - 3$ Myr that it takes a star cluster to form, the total momentum per unit stellar mass $V_{\rm rad} = \dot{V}_{\rm rad} t_{\rm form}$ injected by the radiation field can be competitive with or even exceed that of the outflows.

In the immediate vicinity of forming massive stars, on scales too small to fully sample the IMF, $\dot{V}_{\rm rad}$ can be a factor of $\sim 10-100$ larger -- an individual massive star can have a light to mass ratio in excess of $10^4$ $\lsun/\msun$, a factor of 10 higher than the IMF average. Conversely, the very steep mass-luminosity relation of stars \red{(roughly $L\propto M^{3.5}$ near 1 $M_\odot$, though flattening at much higher masses)} ensures that radiation pressure feedback is dominated by extremely massive stars. As a result, in clusters smaller than $\sim 10^4$ $\msun$ that do not fully sample the IMF, the light to mass ratio is typically much smaller than the mean of a fully-sampled IMF \citep{cervino04a, da-silva12a}. Radiation pressure is therefore less important compared to outflows, which to first order simply follow the mass. Even the Orion Nebula Cluster, the nearest region of massive star formation to the Sun, has a light-to-mass ratio well below the expected value for a fully-sampled IMF \citep{kennicutt12a}. Thus radiation feedback from stars is likely to play a crucial role in the formation of individual high mass stars, in the formation of massive clusters, and possibly even on galactic scales, but is likely to be unimportant in comparison to outflows in low-mass star clusters such as those closest to the Sun.

\red{Radiation pressure begins to become significant for a population of stars once the light-to-mass ratio exceeds $\sim 1000$ $L_\odot/M_\odot$, which corresponds to a mass of $\sim 20$ $M_\odot$ for a single star \citep{krumholz09c}, and $\sim 10^{3.5}$ $M_\odot$ for a star cluster that samples the IMF \citep{cervino04a, KrumThom2012, krumholz13a}. For such stars and star clusters,} much of the luminosity will come in the form of ionizing photons, and thus radiation pressure and photoionization feedback will act together; we defer a more general discussion of the latter process to section \ref{ssec:photoionization}. While most classical treatments of H~\textsc{ii} regions have ignored the effects of radiation pressure, recent analytic models by \citet{krumholz09d}, \citet{krumholz10b}, \citet{fall10a}, and \citet{Murrayetal2010, murray11a} have begun to include it, as did earlier models of starburst galaxies \citep{Thompsonetal2005}. Their general approach is to solve a simple ordinary differential equation for the rate of momentum change of the thin shell bounding an evolving H~\textsc{ii} region due to both gas and radiation pressure.

In this treatment the authors introduce a factor $f_{\rm trap}$ (called $\tau_{\rm IR}$ in the \citeauthor{Murrayetal2010}~models) to account for the boosting of direct radiation pressure force by radiation energy trapped in the expanding shell. The momentum per unit time per unit stellar mass delivered by the radiation field to the gas is $f_{\rm trap} \dot{V}_{\rm rad}$; thus if $f_{\rm trap} \gg 1$, then radiation pressure can be the dominant feedback mechanism almost anywhere massive stars are present. As discussed by \citet{KrumThom2012, krumholz13a}, this factor crudely interpolates between a flow driven purely by the momentum of the radiation field and one that is partly driven by a build-up of radiation energy, in analogy with the ``\red{explosive}" case introduced in Section \ref{ssec:taxonomy}.  Values of $f_{\rm trap} \gg 1$ occur if each photon undergoes many interactions before escaping, while $f_{\rm trap}\sim 1$ corresponds to each stellar photon being absorbed only once, depositing its momentum, and then escaping. In spherical symmetry, it is straightforward to calculate $f_{\rm trap}$ by solving the 1D equation of radiative transfer or some approximation to it (e.g.~the diffusion approximation). Several authors have done this over the years, both analytically and numerically, and found $f_{\rm trap} \gg 1$ \citep[e.g.][]{Kahn1974, YorKru1977, WolCas1986, WolCas1987}. Once one drops the assumption of spherical symmetry, however, the problem becomes vastly more complicated. As a result, the actual value of $f_{\rm trap}$ has been subject to considerable debate both observationally and theoretically, as we discuss below.
 
\subsubsection{Observations of Radiation Pressure Effects}   

Only a few observations to date have investigated the importance of radiation pressure feedback.  \citet{Scovilleetal2001} studied the central regions of M51 and found that radiation pressure from young clusters forming there exceeds their self-gravity. They proposed that this sets an upper limit on cluster masses of $\sim 1000$ $\msun$.  More recent work has studied the giant H~\textsc{ii} region 30 Doradus in the LMC \citep[Figure \ref{fig:30dor}]{lopez11a, pellegrini11a}, as well as a larger sample of H~\textsc{ii} regions in the Magellanic Clouds \citep{lopez13a}. The \citet{lopez11a}~study finds that $f_{\rm trap}$ is generally small, but that nonetheless radiation pressure dominates within 75 pc of the R136 cluster at the center of 30 Doradus. In contrast, \citet{pellegrini11a} argue that radiation pressure is nowhere important in 30 Doradus. (They do not consider the pressure associated with any trapped infrared radiation field, and thus do not address the \red{question} of $f_{\rm trap}$.) This discrepancy is more a matter of definitions than of physics. \citet{lopez11a} adopt the formal definition of radiation pressure as simply the trace of the radiation pressure tensor, while \citet{pellegrini11a} attempt to compute the actual force exerted on matter by radiation. These two definitions produce very different results in the optically-thin interior of 30 Doradus, since in a transparent medium the force experienced by the matter can be small even if the radiation pressure exceeds the gas pressure by orders of magnitude. Regardless of this difference in definition, both sets of authors agree that, in its present configuration, warm ionized gas pressure exceeds radiation pressure at the edge of the swept up shell of material that bounds 30 Doradus. The two studies differ, however, on how this compares to the pressure of shock-heated gas, a topic we defer to Section \ref{ssec:winds}.

\subsubsection{Simulations of Radiation Pressure Feedback}   
 
\begin{figure}
\plotone{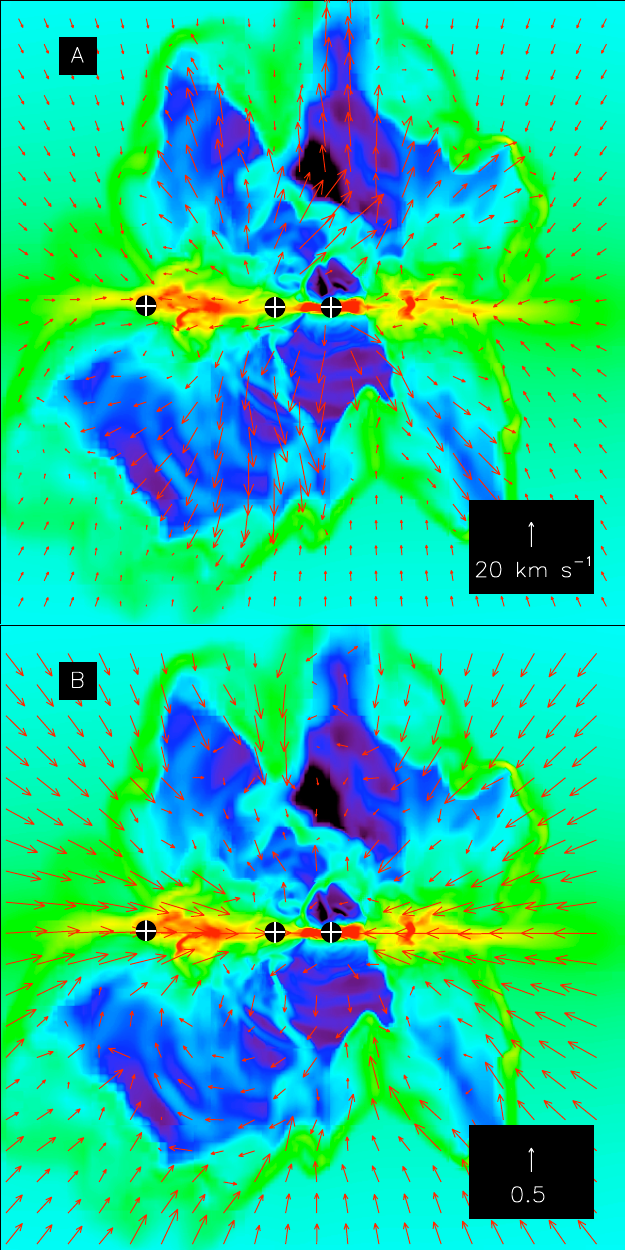}
\caption{\small
\label{fig:krumholz09}
Slices through a simulation of the formation of a 70 $\msun$ binary system, taken from \citet{krumholz09c}. Both panels show a region 6000 AU on a side; colors show volume density on a scale from $10^{-20} - 10^{-14}$ g cm$^{-3}$, and plus signs show stars. In the upper panel, arrows show the velocity, while in the lower panel they show the net radiation plus gravity force $\mathbf{f}_{\rm rad}+\mathbf{f}_{\rm grav}$, with the arrow direction indicating the force direction, and the arrow length scales by $|\mathbf{f}_{\rm rad}+\mathbf{f}_{\rm grav}|/|\mathbf{f}_{\rm grav}|$. \red{These slices show the simulation at a time of $1.0$ mean-density free-fall times, at which point the total stellar mass is $\approx 60$ $\msun$ and the mass of the primary is $\approx 36$ $M_\odot$.}
}
\end{figure}

There has been much more work on simulations of the effects of radiation pressure feedback. On the scales of the formation of individual stars (see the chapter by \textit{Tan et al.}~for more details), \citet{yorke02a} performed two-dimensional radiation-hydrodynamic simulations, and \citet{krumholz05a, krumholz07a, krumholz09c, Krumholzetal2010} performed three-dimensional ones. The general picture established by these simulations, illustrated in Figure \ref{fig:krumholz09}, is that, despite radiation force formally being stronger than gravity on the small scales studied, radiation pressure nevertheless fails to halt accretion.  Gravitational and Rayleigh Taylor (RT) instabilities that develop in the surrounding gas channel the gas onto the star system through non-axisymmetric disks and filaments that self-shield against radiation while allowing radiation to escape through optically thin bubbles in the RT-unstable flow. The radiation-RT instability has been formally analyzed, and linear growth rates calculated, by \citet{jacquet11a} and \citet{jiang13a}.

Some details of this picture have recently been challenged by \citet{kuiperetal2011,kuiperetal2012}, who use a more sophisticated radiative transfer method than \citet{krumholz09c}. Their improved treatment of the direct stellar radiation field increases the rate at which matter is driven away from the star, and as a result the radiation-RT instability does not have time to set in before matter is expelled. While this claim seems very likely to be true as applied to the idealized simulations performed by \citeauthor{kuiperetal2012}, it is unclear whether it would apply in the more realistic case of a turbulent or magnetized protostellar core \citep{myers13a}, or one with a disk that is gravitationally unstable as found in the \citeauthor{krumholz07a}~simulations. In these situations the initial ``seeds" for the instability will be much larger, and the growth to non-linear scale presumably faster. Unfortunately \citeauthor{kuiperetal2012}'s numerical method is limited to treating the case of a single star fixed at the origin of the computational grid, and thus cannot simulate turbulent flows or provide a realistic treatment of the gravitational instabilities expected in massive star disks, which involve displacement of the star from the center of mass \citep{kratter10a}, or fragmentation of the disk into multiple stars \citep{kratter06a, kratter08a, krumholz07a, 2010ApJ...711.1017P,2010ApJ...719..831P, Petersetal2011}. In reality the issue is likely moot in any event, as \citet{krumholz05_outflows} show analytically and \citet{Cunninghametal2011} numerically that protostellar outflows should punch holes through which radiation can escape independent of whether radiation-RT instability occurs or not.

Much less work has been done on larger scales.
\red{A number of authors have introduced subgrid models for radiation pressure feedback, along with other forms of feedback \citep{Hopkinsetal2011, Hopkinsetal2012, hopkins12a, genel12a, agertz13a, aumer13a, stinson13a, bournaud14a}. Others explicitly solve the equation of radiative transfer along rays  emanating from stellar sources, but do not make any attempt to account for radiation that is absorbed and then re-emitted \citep{wise12b, kim13a, kim13b}. However, none of these simulations include a fully-self-consistent treatment of the interaction of the radiation field with the ISM, and as result they are forced to adopt a value of $f_{\rm trap}$, either explicitly or implicitly. The outcome of the simulations depends strongly on this choice. At one extreme, some authors adopt values of $f_{\rm trap} \gg 1$, in some cases $f_{\rm trap} \sim 50$ \citep[e.g.,][]{Hopkinsetal2011, aumer13a}, and find that radiation pressure is a dominant regulator of star formation in rapidly star-forming galaxies. At the other extreme, models with more modest values of $f_{\rm trap}$ find correspondingly modest effects \citep[e.g.,][]{kim13b, agertz13a}. The only large-scale fully-radiation-hydrodynamic simulations published thus far are those of \citet{KrumThom2012, krumholz13a}, who find that real galaxies on large scales likely have $f_{\rm trap}\sim 1$,
} because radiation-RT instability reduces the efficiency of radiation-matter coupling far below the value for a laminar radiation distribution. (\citet{SocratesSironi2013} have also argued \red{against values of $f_{\rm trap} \gg 1$}, for somewhat different reasons.) There is some potential worry about the treatment of the direct radiation field in the \citeauthor{KrumThom2012} models, following the points made by \citet{kuiperetal2012}. \citeauthor{KrumThom2012}~argue that if $f_{\rm trap} \gg 1$ then the direct radiation pressure force is by definition unimportant, and thus that their treatment of radiation is adequate for investigating whether $f_{\rm trap}$ is indeed large. The simulations of \citet{jiang13a}, which use a more sophisticated treatment of radiation pressure than either \citeauthor{KrumThom2012} or \citeauthor{kuiperetal2012}, are qualitatively consistent with this conclusion. However, there is clearly a need for further numerical investigations with more accurate radiation-hydrodynamic methods to fully pin down the correct value of $f_{\rm trap}$ for use in subgrid models.

\bigskip
\section{\textbf{\red{EXPLOSIVE} FEEDBACK}}
\bigskip

\subsection{\textbf{Main Sequence Winds from Hot Stars}}
\label{ssec:winds}
\bigskip

\subsubsection{Budget and Relative Importance}

While all stars that accrete from disks appear to produce protostellar outflows, only those with surface temperatures above $\sim 2.5\times 10^4$ K produce strong winds \citep{vink00a}. Main sequence stars reach this temperature at a mass of $\sim 40$ $\msun$, and stars this massive have such short Kelvin-Helmholtz times that, even for very high accretion rates, they reach their main sequence surface temperatures while still forming \citep{hosokawa09a}. As a result, hot stellar winds will begin to appear very early in the star formation process. These winds carry slightly less momentum than the stellar radiation field \citep{kudritzki99a}; a calculation using Starburst99 \citep{leitherer99a, vazquez05a} gives $\red{\dot{V}}_{\rm msw}=9$ km s$^{-1}$ Myr$^{-1}$ per unit mass of star formed for a zero-age population.

\red{This estimate is based on non-rotating stars of Solar metallicity, using the wind prescriptions of \citet{leitherer92a} and \citet{vink00a}. Stars of lower metallicity will have significantly lower wind momentum fluxes, due to the reduced efficiency of line-driving in a stellar atmosphere containing fewer heavy elements \citep{vink01a}. Conversely, stellar rotation can increase the instantaneous momentum flux by a factor of a few, and the integrated momentum over the stellar lifetime by a factor of $\sim 10$, with significant uncertainty arising from the poorly-known distribution of birth rotation rates \citep{maeder00a, maeder10a, ekstrom12a}. There is also significant uncertainty on the momentum budget at ages greater than a few Myr, stemming from our poor knowledge of how exactly massive stars evolve into luminous blue variables, red and yellow supergiants, and Wolf-Rayet stars.}

\red{Despite these uncertainties, even the highest plausible stellar wind momentum estimates yield injection rates at most comparable to the stellar radiation field.} Thus if stellar winds represent a momentum-driven form of feedback, they should only provide a mild enhancement of radiation pressure. However, hot star winds can have terminal velocities of several times $10^3$ km s$^{-1}$ \citep[e.g.][]{Castor75b, leitherer92a}, so when they shock against one another or the surrounding ISM, the post-shock temperature can exceed $\sim 10^7$ K. At this temperature radiative cooling times are long \citep{castor75a, weaver77a}, so shocked stellar winds might build up an energy-driven, adiabatic flow that would make them far more effective than radiation. On the other hand, they might also leak out of their confining shells of dense interstellar matter, which would greatly reduce the pressure build-up and lead to something closer to the momentum-limited case.

Whether shocked stellar wind gas does actually build up an energy-driven flow and thereby become an important feedback mechanism has been subject to considerable debate\red{, and we discuss the available observational and theoretical evidence in the next section.} To frame the discussion, consider an H~\textsc{ii} region with a volume $V$ and a pressure at its outer edge $P$, within which some volume $V_w$ is occupied by X-ray emitting post-shock wind gas at pressure $P_w$. \citet{yeh12a} introduce the wind parameter
\begin{equation}
\Omega \equiv \frac{P_w V_w}{PV - P_w V_w}
\end{equation}
as a measure of the relative strength of winds. The virial theorem implies that $PV$ is what controls the large-scale dynamics, so $\Omega\gg 1$ (as expected for models such as those of \citealt{castor75a} and \citealt{weaver77a}) indicates that the large-scale dynamics are determined primarily by winds, while $\Omega \ll 1$ (as expected in models where the post-shock wind gas undergoes free expansion, e.g., \citealt{chevalier85a}) indicates they are unimportant. Note that $P$ and $V$ include any form of feedback that contributes pressure to and occupies volume within the H~\textsc{ii} region, not just the pressure and volume associated with the $\sim 10^4$ K photoionized gas (see below). Thus $\Omega$ should be thought of as measuring the contribution of stellar winds to the total dynamical budget. \red{As we discuss in the next section, the true value of $\Omega$ remains an open question in both observations and theory.}

\subsubsection{Observations and Theory}

Observations of stellar wind feedback were revolutionized by the launch of the \textit{Chandra X-Ray Observatory}, which for the first time made it possible to detect X-ray emission from the hot post-shock wind gas in H~\textsc{ii} regions \citep{townsley03a}. The sample of H~\textsc{ii} regions with X-ray measurements includes M17 and the Rosette Nebula \citep{townsley03a}, the Carina Nebula \citep{townsley11a}, the Tarantula Nebula / 30 Doradus \citep{townsley06a, lopez11a, pellegrini11a}, and a few tens of smaller H~\textsc{ii} regions in the Magellanic Clouds \citep{lopez13a}. A measurement of the X-ray luminosity and spectrum can be used to infer $P_w$, at least up to an unknown volume filling factor. The observations conducted to date strongly rule out the largest predicted values of $\Omega$, but the exact value is still debated. In 30 Doradus, probably the best-studied case, \citet{lopez11a} and \citet{pellegrini11a} report similar estimates for the pressures of $10^4$ K photoionized gas and radiation, but \citeauthor{pellegrini11a}'s estimate of the X-ray-emitting gas pressure is $\sim 2$ orders of magnitude larger. Most of this discrepancy is due to differing assumptions about the volume filling factor of the emitting gas, with \citeauthor{lopez11a}~assuming it is of order unity and \citeauthor{pellegrini11a}~arguing for a much smaller value, which would imply higher $P_w$ but also lower $V_w$. \citeauthor{lopez11a}'s reported values give $\Omega \ll 1$, while the value of $\Omega$ based on \citeauthor{pellegrini11a}'s modeling is unclear because they do not report values for $V_w$. However, they likely obtain $\Omega\ll 1$ too, since, all other things being equal, a reduction in the filling factor tends to lower $V_w$ more than it raises $P_w$.

While X-rays are the most direct way of constraining $\Omega$, some optical and infrared line ratios are sensitive to it as well \citep{yeh12a, yeh13a, verdolini13a}. There are significant modeling uncertainties associated with the assumed geometry, but these are quite different from the filling factor issues that hamper X-ray measurements. \citet{yeh12a} find that available data favor $\Omega < 1$, but this is a preliminary analysis.

\begin{figure}[ht]
\plotone{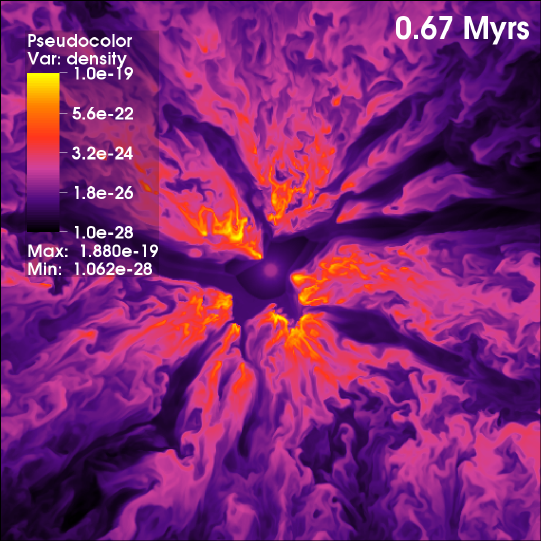}
\caption{\small
\label{fig:rogers_pittard13}
Slice through a simulation of a molecular clump with hot stellar winds launched by a central star cluster. The region shown is 32 pc on a side at a time $0.67$ My after wind launching begins. The color shows the density; the low-density channels shown in black are filled with hot, escaping wind material. Adapted from Figure 3 of \citet{rogers13a}.
}
\end{figure}

Several authors have also made theoretical models of wind feedback. \red{Much of this work has treated the ISM surrounding the star as a simple uniform-density medium, and has focused on instead on the circumstellar medium, within which there can be complex interactions between components of the wind launched at different stages of stellar evolution \citep{garcia-segura96b, garcia-segura96c, freyer03a, freyer06a, toala11a}. This work, while clearly important for the study of circumstellar bubbles, offers limited insight into how wind feedback affects the process of star formation. Similarly,}
building on the classical \citet{castor75a} and \citet{weaver77a} models, a number of authors have made increasingly-sophisticated spherically-symmetric models of stellar wind bubbles, including better treatments of conduction and radiative cooling \citep{capriotti01a, tenorio-tagle07a, arthur12a, silich13a}. They find that stellar wind feedback can be dominant, but since these models necessarily exclude leakage, it is unclear how much weight to give to this conclusion.

\red{Studies that include multi-dimensional effects in a complex, star-forming interstellar medium are significantly fewer.} \citet{harper-clark09a} present analytic models for cold shells driven by a combination of ionized gas pressure, radiation pressure, and wind pressure, including parameterized treatments of wind leakage. These models are able to fit the observations by adopting fairly strong leakage of hot gas, and the required values of the leakage parameter suggest that hot gas is subdominant compared to other forms of feedback. Two-dimensional simulations by \citet{tenorio-tagle07a} and three-dimensional ones by \citet{dale08a} and \citet{rogers13a} generally find that leakage is a very significant effect, as illustrated in Figure \ref{fig:rogers_pittard13}, with a majority of the injected wind energy escaping rather than being used to do work on the cold ISM. However, even with these losses, the multi-dimensional grid-based simulations do show that a wind of hot gas is eventually able to entrain the cold ISM via Kelvin-Helmholtz instabilities and eventually remove all the cold gas from a forming star cluster. (Simulations using \red{older} formulations of SPH cannot capture this effect due to their difficulties in modeling the Kelvin-Helmholtz instability -- see \citealt{agertz07a}. \red{Newer SPH methods can overcome this limitation \citep[e.g.,][]{price08a, read12a, hopkins13f}, but all SPH simulations of stellar wind feedback published to date use the older methods.}) These simulations, however, do not include radiation pressure or other forms of feedback, and it is unclear if winds would be dominant in competition with other mechanisms.
 
In addition, all the multidimensional simulations performed to date lack the resolution and the sophisticated microphysics required to handle a number of other potentially important effects. For example, the development of a turbulent interface between the cold and hot gas might greatly enhance the rate of conduction, thus lowering the temperature in the hot gas to $\sim 10^5$ K so that radiative losses via far-UV metal lines become rapid \citep{mckee84a}. Another possibly important effect  is mixing of dust grains into the hot gas, where, until they are destroyed by sputtering, they can remove energy via collisional heating followed by thermal radiation. In light of the continuing controversy over the importance of stellar winds, reinvestigation of these topics using modern hydrodynamic techniques is urgently needed.

\bigskip
\subsection{\textbf{Photoionization Feedback}}
\label{ssec:photoionization}
\bigskip

Stars with masses $\gtrsim10$ $\msun$ emit very large quantities of ionizing photons, creating ionized bubbles -- H~\textsc{ii} regions. Equilibrium between heating and cooling processes inside H~\textsc{ii} regions results in them having remarkably constant temperatures of $\approx 10^{4}$ K and internal sound speeds of $\approx10$ km s$^{-1}$ \citep{2006agna.book.....O}. The overpressure in the bubble causes it to expand at velocities of order the sound speed. In a uniform medium, this leads to the well known Spitzer solution \citep{1978ppim.book.....S}. As is the case with stellar winds, the interaction between the cold molecular gas and the over-pressured, expanding hot gas is physically complex, and thus it is not trivial to assign a momentum budget to it and compare it to other sources of feedback. We will return to this topic below.

Observationally,  H~\textsc{ii} regions are extremely bright at radio wavelengths (due to radio recombination lines and bremsstrahlung) and in the infrared (due to reprocessing of stellar radiation by dust), making it possible to study them over large distances. H~\textsc{ii} regions are often divided into (ultra-- or hyper--) compact and diffuse types, and this was originally thought to be an evolutionary sequence resulting from expansion. However, observations by, e.g., \citet{1989ApJS...69..831W} and \citet{1994ApJS...91..659K} revealed that UCH~\textsc{ii} regions rarely resemble classical Str\"omgren spheres. Common morphologies are cometary, core-halo or shell-like, and irregular. \red{Simulations suggest that these morphologies result from variations in the mass distribution in the immediate vicinity of the ionizing stars, and are likely to be variable over $\sim$kyr or even shorter timescales \citep{2010ApJ...711.1017P}.} \cite{1999ApJ...514..232K} and \cite{2001ApJ...549..979K} found that many compact H~\textsc{ii} regions are embedded in larger diffuse ionized regions, leading \cite{2003ApJ...596..362K} to propose that UCH~\textsc{ii} regions are dense cores embedded in champagne flows. More recent observational work has concentrated on the interaction of H~\textsc{ii} regions with infrared dust bubbles \citep[e.g][]{2008ApJ...681.1341W,2010A&A...523A...6D,2011ApJS..194...32A}, molecular gas \citep[e.g.][]{2011ApJS..194...32A} and stellar winds \citep[e.g][]{townsley03a}.

On small scales, photoionization is sometimes suggested to limit the growth of OB stars. However, \cite{1995RMxAC...1..137W} showed that the expansion of an H~\textsc{ii} region could be stalled or even reversed by an accretion flow. \cite{2003ApJ...599.1196K} generalized this result by showing that accretion onto an ionizing star can proceed through a gravitationally-trapped H~\textsc{ii} region; observations of ionized accretion flows  support this picture \citep{2006ApJ...637..850K, 2007ApJ...663.1092K}. Simulations by \citet{2010ApJ...711.1017P,2010ApJ...719..831P} also found that ionizing sources were unable to disrupt accretion flows onto them until material was drained from the flows by other, lower-mass accretors. Moreover, \cite{2010ApJ...721..478H,2012ApJ...760L..37H} point out that accretion onto massive stars at high but not unreasonable rates causes the stars to expand, cooling their photospheres and reducing their ionizing fluxes, further easing accretion. \citet{2012ApJ...758..137K} model the consequences of accretion--induced expansion and conclude that the shrinking of the H~\textsc{ii} region due to the drop in ionizing flux may be observable with facilities such as EVLA or ALMA.
Statistical analysis of the correlation between the bolometric luminosities of massive young stellar objects and the ionizing photon fluxes required to drive the H~\textsc{ii} regions around them provides direct evidence for this effect \citep{mottram11a, davies11a}.

At larger scales, there are three major outstanding questions regarding H~\textsc{ii} regions. The first -- whether they are able to trigger star formation -- is discussed in Section \ref{ssec:triggering}. The second is whether H~\textsc{ii} region expansion is able to drive GMC turbulence. There have been several simulations of H~\textsc{ii} region expansion in turbulent clouds \citep[e.g.,][]{2006ApJ...647..397M,2007ApJ...668..980M,2012A&A...546A..33T,2012MNRAS.424..377D,2013MNRAS.430..234D} but few authors have addressed this issue in detail. \red{\cite{2006ApJ...647..397M} simulated the expansion of an H~\textsc{ii} region in a turbulent cloud and found that substantial kinetic energy was deposited in the neutral gas, although they did not show explicitly that this actually kept the cold gas turbulent in the sense of maintaining a self--similar velocity field over some range of scales. In their simulations of an ionizing source inside a fractal cloud, \cite{2012MNRAS.427..625W} also showed that the kinetic energy of the cold gas was strongly influenced by ionization -- more so than by gravity -- and that a large fraction of this energy resided in random motions, which they identified as turbulence.}

\red{\citet{2009ApJ...694L..26G} simulated plane--parallel photoionization of a turbulent box and analyzed the power--spectra of the velocity field both with and without the influence of feedback. They found that ionization was an efficient driver of turbulence, although with a substantially flatter power--spectrum than the Kolmogorov velocity field with which the box was seeded. However, because they were irradiating the whole of one side of their simulation domain, they were driving turbulence on the largest scale available and it is not clear that this result applies more generally. In the case of pointlike ionizing sources inside a large cloud, the H~\textsc{ii} regions must grow to fill a large fraction of the system volume in order for turbulent driving to be effective on the largest scales and for the turbulent cascade to operate. Semi--analytic models of GMCs including H~\textsc{ii} region feedback by \cite{KruMatMcK2006} and \cite{goldbaum11a}, and simulations of H~\textsc{ii} regions evolving in isolated clouds by \cite{2012MNRAS.427..625W}, \cite{2012MNRAS.424..377D} and \cite{2013MNRAS.430..234D}, suggest that is often likely to be the case, consistent with observations that many H~\textsc{ii} regions are champagne flows.}

\begin{figure}
\plotone{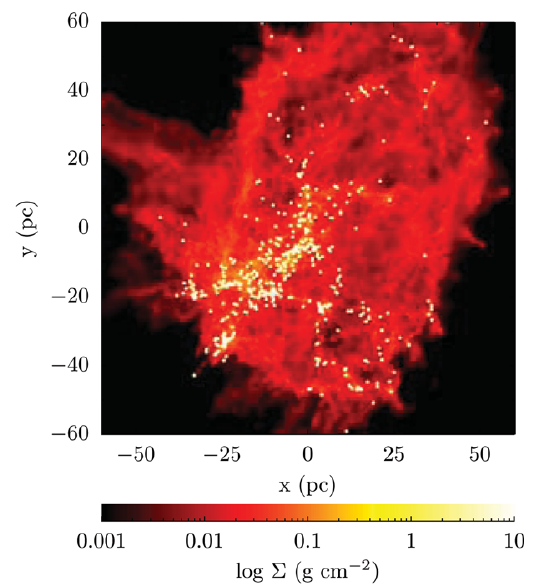}
\plotone{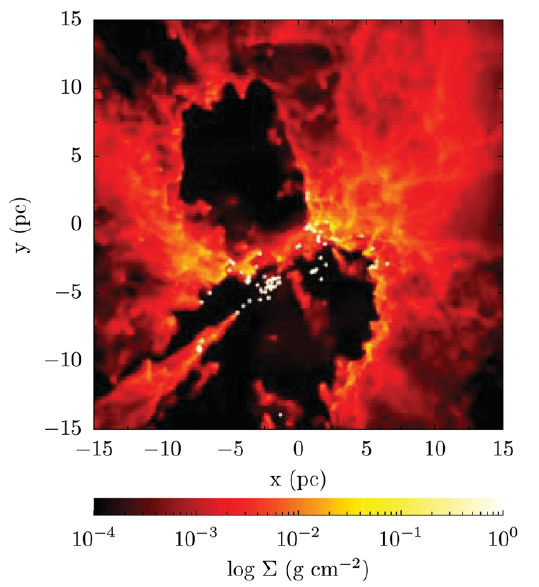}
\caption{\small
\label{fig:dale12}
Column density maps from two simulations of photoionization feedback. The upper panel shows a cloud with an escape speed $>10$ km s$^{-1}$, within which photoionization has done little to inhibit star formation or remove mass. The lower panel shows a cloud with an escape speed $<10$ km s$^{-1}$, where a majority of the mass has been ejected by photoionization feedback. Adapted from \citet{2012MNRAS.424..377D}.
}
\end{figure}

The final question is whether H~\textsc{ii} regions can disrupt protoclusters and terminate star formation at low efficiencies. \citet{1979MNRAS.186...59W}, \citet{1990ApJ...349..126F}, \citet{1994ApJ...436..795F}, \citet{Matzner2002}, \citet{KruMatMcK2006}, and \citet{goldbaum11a} found that photoinization should be effective in destroying clouds. However, these authors considered the effects of ionization on smooth clouds. The picture from recent numerical simulations of H~\textsc{ii} regions expanding in highly--structured clouds is less clear. While O stars located on the edges of clouds can drive very destructive champagne flows \citep[e.g.,][]{1979MNRAS.186...59W}, massive stars are usually to be found embedded deep inside clouds. In addition, molecular clouds usually possess complex density fields, and the massive stars are often located inside the densest regions. \citet{2005MNRAS.358..291D} found that the influence of a photoionizing source could be strongly limited by dense large scale structures and accretion flows. \citet{2012MNRAS.427..625W} found that ionization was very destructive to $\sim10^{4}$ $\msun$ fractal clouds on timescales of 1 Myr, but \citet{2011MNRAS.414..321D} and \citet{2012MNRAS.424..377D,2013MNRAS.430..234D} simulated expanding H~\textsc{ii} regions in turbulent clouds with a range of radii ($\sim 1-100$ pc) and masses ($10^{4}-10^{6}$ $\msun$) and found that the influence of ionization depends critically on cloud escape velocities. It is very effective in clouds with escape speeds well below $\sim 10$ km s$^{-1}$, but becomes ineffective once the escape speed reaches this value. Figure \ref{fig:dale12} shows an example. In high escape-speed clouds, radiation pressure may dominate instead -- see Section~\ref{ssec:radpressure}.

A final caution is that, with a few exceptions \citep{2007ApJ...671..518K,2011MNRAS.414.1747A,Petersetal2011,2012ApJ...745..158G} the simulations of H~\textsc{ii} regions performed to date ignore magnetic fields, and the coupling between magnetic fields an ionizing radiation can generate unexpected effects. We discuss this further in Section \ref{sec:interactions}.

\bigskip
\subsection{\textbf{Supernovae}}
\bigskip

\red{Supernovae represent a final source of explosive feedback}. For every 100 $\msun$ of stars formed, $\sim 1$ star will end its life in a type II supernova; the supernovae are distributed roughly uniformly in time from $\approx 4-40$ Myr after the onset of star formation, with a slight peak during the first Myr after the explosions begin \citep[e.g.][]{Matzner2002}. Each supernova yields $\sim 10^{51}$ erg, much of which ends up as thermal energy in a hot phase with a long cooling time. However, as pointed out by \citet{krumholz09d} and \citet{fall10a}, supernovae are of quite limited importance on the scales of individual star clusters simply due to timescale issues. The first supernovae do not occur until roughly 4 Myr after the onset of star formation. In comparison, the crossing time of a protocluster is $t_{\rm cr} = 2R/\sigma$, where $R$ and $\sigma$ are the radius and velocity dispersion, respectively. Since protoclusters have virial ratios $\alpha_{\rm vir} \approx 1$, and form a sequence of roughly constant surface density with $\Sigma\sim 1$ g cm$^{-2}$ \citep{fall10a}, the crossing time varies with protocluster mass as $t_{\rm cr} \approx 0.25 (M/10^4\,M_\odot)^{1/4}\mbox{ Myr}$ \citep{tan06a}. Thus protoclusters have crossing times smaller than the time required for supernovae to start occurring unless their masses are $\ga 10^6$ $\msun$. In the absence of feedback, protocluster gas clouds convert the great majority of their gas to stars in roughly a crossing time, so supernovae cannot be the dominant form of feedback in clusters smaller than $\sim 10^6$ $\msun$ -- such clusters would convert all of their mass to stars before the first supernova occurred, unless some other mechanism were able to delay star formation for several crossing times and allow time for supernovae to begin.

This conclusion is consistent with observations of the most massive clusters that host stars that will end their lives as supernovae. In 30 Doradus there is only one detectable supernova remnant (the bright blue spot in the lower right corner of Figure \ref{fig:30dor}; \citealt{lopez11a}), and its radius is far smaller than that of the evacuated bubble. Similarly, in Westerlund 1, which has also ejected its central gas, there has been a supernova \citep{muno06b}, but no corresponding supernova remnant has been detected \citep{muno06a}. This is likely because the gas had already been expelled before the supernova occurred, and thus the ejecta have yet to encounter material dense enough to produce on observable shock. However, we emphasize that the conclusion that supernovae are unimportant for clusters does not apply on the larger scales of diffuse giant molecular clouds or galaxies, which have crossing times that are comparable to or significantly larger than the lifetime of a massive star.

\bigskip
\section{\textbf{THERMAL FEEDBACK}}
\bigskip

Although radiative transfer was included in some of the very earliest calculations of star formation \citep[e.g.][]{Larson1969}, the importance of thermal feedback for star formation has only been recognized recently. This is true even for the case of massive star formation, where for a long time in the literature ``radiative feedback" meant radiation pressure (see Section \ref{ssec:radpressure}), not radiative heating. This began to change in the 1980s, when three-dimensional calculations of low-mass star formation began to include radiative transfer using the Eddington approximation \citep{Boss1983, Boss1984, Boss1986}.  Such calculations showed that perturbed molecular cloud cores containing several Jeans masses initially readily fragmented to form binary systems and that dynamical collapse and fragmentation is terminated by the thermal heating after the clouds become opaque \citep[the so-called opacity limit for fragmentation;][]{LowLyn1976,Rees1976}.  However, over the past few years it has been realized that thermal feedback could play a much greater role in star formation than simply setting the minimum fragment mass.

\bigskip
\noindent
\subsection{\textbf{Origin of thermal feedback}}
\bigskip
\label{ssec:feedbackorigins}

Thermal feedback is inevitable in the star formation process as gravitational potential energy is converted to kinetic and thermal energy during collapse.  Initially, the rate of compression of the gas is low and the additional thermal energy is quickly radiated, resulting in an almost isothermal collapse \citep{Larson1969}.  However, as the rate of collapse increases, compressional heating eventually exceeds the radiative losses \citep{MasInu1999} and the collapse at the center of the cloud transitions to an almost adiabatic phase leading to the formation of a pressure-supported object with a mass of a few Jupiter masses and a radius of $\sim 5$~AU \citep[the so-called first hydrostatic core;][]{Larson1969}.  The first core accretes through a supercritical shock from which most of the accretion luminosity is radiated away \citep{tomida10b, tomida10a, Commerconetal2011b, schonke11a}. This results in heating of the surrounding gas which, although modest, may affect fragmentation \citep{Bossetal2000, WhiBat2006}.

Once the center of the first hydrostatic core reaches $\approx 2000$~K, the dissociation of molecular hydrogen initiates a second dynamical collapse resulting in the formation of the second, or stellar, core \citep{Larson1969, MasInu2000}.  The stellar core forms with a few Jupiter masses of gas and a radius of $\sim 2$ $\rsun$.  Because gravitational potential energy scales inversely with radius, the formation of the stellar core is associated with a dramatic increase in the luminosity of the protostar and significant heating to distances of hundreds of AU from the stellar core \citep[e.g.][]{WhiBat2006}.  Recent radiation hydrodynamical simulations of stellar core formation have shown that this burst of thermal feedback, due to accretion rates of $\sim 10^{-3}$ $\msun$ yr$^{-1}$ which last for a few years, can be great enough to launch pressure-driven bipolar outflows (in the absence of magnetic fields) as gas and dust heated by the accretion luminosity expands and bursts out of the first hydrostatic core in which the stellar core is embedded \citep{Bate2010, Bate2011, schonke11a}.

Once a stellar core forms, there are three sources of thermal feedback: radiation originating from the core itself, luminosity from accretion onto the star, and luminosity from continued collapse of the cloud and disk accretion. For low-mass protostars ($\lsim 3$ $\msun$) accreting at rates $\gsim 10^{-6}$ $\msun$ yr$^{-1}$, accretion luminosity dominates both the intrinsic stellar luminosity \citep[e.g.][]{palla91a, palla92a, hosokawa09a} and the luminosity of the larger scale collapse \citep{offner09a, bate12a}.  For example, the accretion luminosity 
\begin{equation}
\label{eq:acclum}
L_{\rm acc} \approx  \frac{G M_* \dot{M}_*}{R_*}.
\end{equation}
of a star of mass $M_*=1$ $\msun$ with a radius of $R_* =2\rsun$ \citep[e.g.,][]{hosokawa09a} accreting at $\dot{M}_* = 1\times 10^{-6}$ $\msun$ yr$^{-1}$ is $\approx 15$ $\lsun$ whereas the luminosity of the stellar object itself is $\approx 1$ $\lsun$. For intermediate-mass protostars ($M_*>3-9$ $\msun$), whether the  accretion luminosity or the intrinsic luminosity dominates depends on accretion rate, while for masses greater than $\approx 9$ $\msun$ the intrinsic stellar luminosity dominates for all reasonable accretion rates ($\lsim 10^{-3}$ $\msun$ yr$^{-1}$).  

Several authors have considered the impact of accretion luminosity on the temperature distribution in protostellar cores of a variety of masses, both analytically and numerically \citep{chakrabarti05a, Krumholz2006, robitaille06a, robitaille07a}. These models show that even sub-solar-mass protostars could heat the interior of cores to temperatures in excess of 100~K to distances $\sim 100$~AU or 30~K to distances $\sim 1000$~AU. \citet{Krumholz2006} points out that this could significantly inhibit fragmentation of massive cores to form stellar groups and multiple star formation in low-mass cores.
 
Due to the inverse radial dependence of equation \ref{eq:acclum}, the luminosity from accretion onto a star will generally dominate that produced by either the accretion disk or continued collapse on larger scales.  However, there are a large number of uncertainties that make accurate determination of the luminosity difficult.  The evolution of the stellar core depends both on its initial structure at formation and on how much energy is advected into the star as it accretes \citep{HarCasKen1997,TouLivBon1999,BarChaGal2009, HosOffKru2011}.  Different assumptions give different intrinsic luminosities and stellar radii.  The latter uncertainty translates into an uncertainty in the accretion luminosity.  An unknown fraction of the energy will also drive jets and outflows rather than being emitted as accretion luminosity.  Furthermore, protostars may accrete much of their mass in bursts (see the chapter by \textit{Audard et al.}). If this is the case, protostars may spend the majority of their time in a low-luminosity state with only brief periods of high luminosity.  Thus, \red{\citet{StaWhiHub2011, stamatellos12a}} recently argued that numerical calculations assuming continuous radiative feedback may overestimate its effects.  See \cite{bate12a} for a detailed discussion of these issues, and numerical issues related to modeling protostars with sink particles.

\bigskip
\noindent
\subsection{\textbf{Observations of thermal feedback}}
\bigskip
\label{ssec:thermal_observations}

Observations provide strong evidence for thermal feedback. The first indirect hints came from observations of a narrow CO($6-5$) component around low-mass YSOs, which models suggested was produced by a $\sim 1000$ AU-scale region heated to $\sim 100$ K by UV photons interacting with the walls of an outflow cavity \citep{spaans95a}. More recently, several new telescopes and instruments have allowed us to obtain much larger samples to study the effects of thermal feedback on sub-parsec scales.  Combining information on the thermal structure, thermodynamic properties and fragmentation of cluster forming clumps can potentially constrain the importance of thermal feedback on cluster formation \citep{zhang2009,zhang2011,Longmoreetal2011,wang2011}.

In the vicinity of low-mass stars, \citet{van-kempen09a, van-kempen09b} used APEX-CHAMP$^+$ to obtain spatially-resolved maps of high-$J$ CO lines that trace warm gas around $\sim 30$ nearby sources; \citet{van-kempen10a} complement this with \textit{Herschel}/PACS spectroscopy to study the spectra in detail. \citet{visser12a} and \citet{yildiz12a} model the data and confirm that they are consistent with heating by a combination of stellar photons and UV produced by shocks when the jet interacts with the circumstellar medium. All these observations point to the conclusion that $\sim 1000$ AU-scale, $\sim 100$ K regions are ubiquitous around the outflow cavities produced by embedded low-mass protostars.

In more massive regions, \citet{Longmoreetal2011} analyzed the density and temperature structure of the massive protocluster G8.68-0.37 with an estimated mass of $\approx 1500$ $\msun$.  Combining Australia Telescope Compact Array (ATCA) and Submillimeter Array (SMA) observations with radiative transfer modeling they found radial temperature profiles $T \propto r^{-0.35}$ with temperatures of $\approx 40$ K at distances of 0.3~pc from the cluster center.  \citet{zhang2009} and \citet{Wangetal2012} used Very Large Array (VLA) ammonia observations of another massive protocluster G28.34+0.06 and found that warmer gas seemed to be associated with outflows, but that the protostars themselves did not seem to provide significant thermal feedback on scales of 0.06~pc. The chemical species observed with \textit{Herschel} also provide evidence for thermal feedback associated with outflows \citep{bruderer10a}.

By using multi-wavelength imaging, temperature maps of star-forming regions can be constructed.  \citet{Hatchelletal2013} used James Clerk Maxwell Telescope (JCMT) SCUBA-2 observations to map the temperature structure in NGC 1333, detecting heating from a nearby B star, other young infrared/optical stars in the cluster, and embedded protostars. Temperatures ranged from 40~K at distances of a few thousand AU from some of the more luminous stars to 20-30~K on scales of $\approx 0.2$~pc in the north of the star-forming region. They argued that heating from existing stars may lead to increased masses to the next generation of stars to be formed in the region.  \citet{SiciliaAguilaretal2013} used multi-wavelength Herschel observations to create temperature maps of the Corona Australis region, also detecting heating from protostars on scales of thousands of AU.

Therefore, from both theory and observation it is now clear that even low-mass protostars produce substantial thermal feedback on the gas and dust surrounding them.  As we will discuss in the next two sections, thermal feedback may be a crucial ingredient in producing the stellar IMF.

\bigskip
\subsection{\textbf{Influence on the IMF: low-mass end}}
\bigskip

About fifteen years ago it became computationally feasible to perform hydrodynamic simulations of the gravitational collapse of molecular clouds to produce groups of protostars \citep[e.g.][]{BonBatClaPri1997,KleBurBat1998,bate03a}.  Early simulations treated the gas either isothermally or using simple barotropic equations of state. These calculations were able to produce IMF-like stellar mass distributions, but with two major problems.  First, simulations systematically over-produced brown dwarfs compared with observed Galactic star-forming regions \citep{bate03a, BatBon2005, Bate2009a}, particularly in the case of decaying turbulence \citep{OffKleMcK2008}.  Second, the characteristic mass of stars formed in the simulations was proportional to the initial Jeans mass \citep{klessen00a, klessen01a, BatBon2005, jappsen05a, BonClaBat2006}, while there is no firm evidence for such environmental dependence in reality \citep{BasCovMey2010}.

To explain why the characteristic stellar mass does not vary strongly with environment, several authors have suggested that it might be set by microphysical processes that cause the equation of state to deviate subtly from isothermal, for example a changeover from cooling being dominated by line emission to being dominated by dust emission at some characteristic density \citep{larson85a, larson05a}. Simulations using such non-isothermal equations of state show that they are capable of producing a characteristic stellar mass that does not depend on the mean density or similar properties of the initial cloud \citep{jappsen05a}. However, these models neglected the effects of stellar radiative feedback, which, as discussed above, heats the gas near existing protostars and inhibits fragmentation. Indeed, radiation-hydrodynamic simulations show that, once radiative feedback is included, the proposed non-isothermal equations of state do not provide a good description of the actual temperature structure \citep{krumholz07a, urban09a}.

The first cluster-scale calculations to include radiative transfer \citep{bate09a, offner09a, urban10a} showed that this drastically reduces the amount of fragmentation even in regions that produce only low mass stars. As a result the typical stellar mass is greater than without thermal feedback, greatly reducing the ratio of brown dwarfs to stars and bringing it into good agreement with the observed Galactic IMF.  However, potentially of even more importance, \cite{bate09a} also showed that radiative feedback apparently removed the dependence of the IMF on the initial Jeans mass of the cloud and, therefore, could be a crucial ingredient for producing an invariant IMF.  \cite{krumholz11e} took \citeauthor{bate09a}'s argument even further and proposed that the characteristic mass of the IMF may be linked, through thermal feedback, to a combination of fundamental constants.

\begin{figure}
\epsscale{0.7}
\plotone{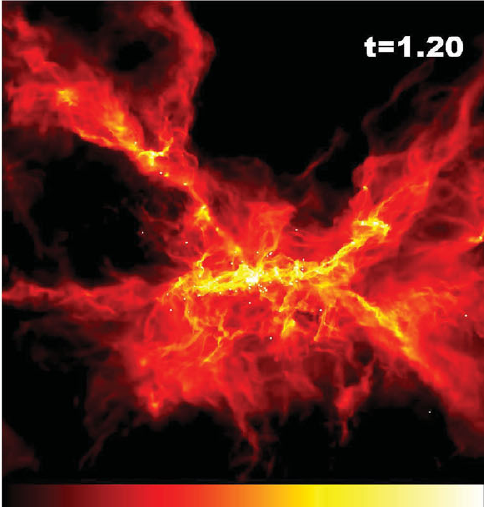}
\plotone{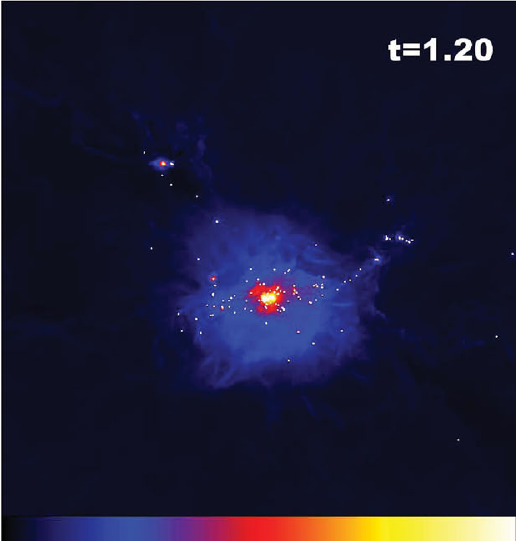}
\plotone{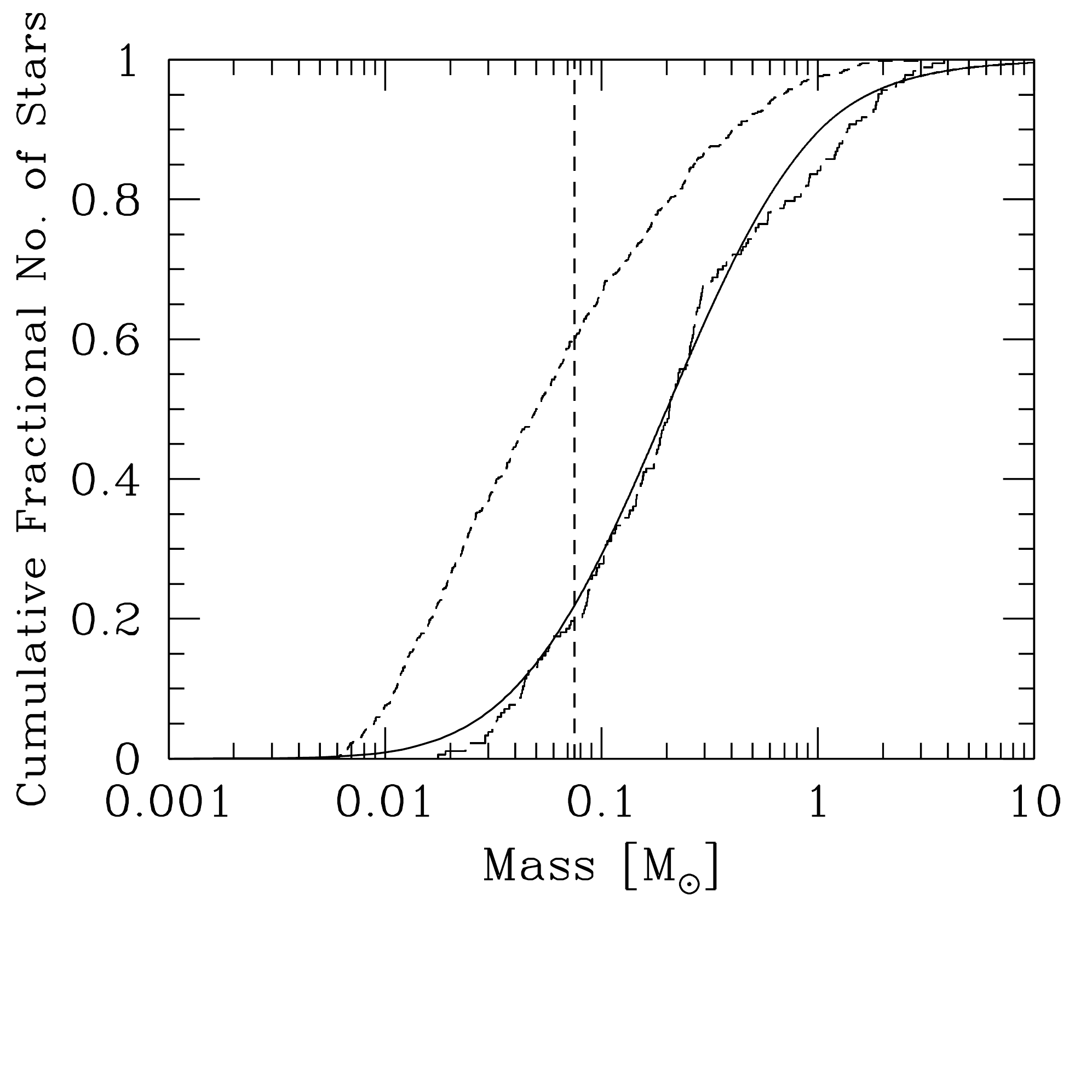}
\epsscale{1.0}
\caption{\small
\label{fig:bate12}
Results from a radiation-hydrodynamic simulation of the formation of a star cluster in a 500 $\msun$ gas cloud by \citet{bate12a}. The top panel shows the gas column density $\Sigma$ at the end of the simulation on a logarithmic scale from $-1.4 < \log\Sigma< 1.0$ with $\Sigma$ in g cm$^{-2}$. The middle panel shows the density-weighted temperature on a linear scale from $9-50$ K. The bottom panel shows the cumulative mass distribution of the stars at the end of the simulation (dashed line), and for comparison the observed IMF \citep[solid line]{chabrier05a} and the mass distribution from a simulation with identical initial conditions and evolved to the same time using a barotropic equation of state \citep[dotted line]{Bate2009a}. Figures adapted from \citet{bate12a}.
}
\end{figure}

More recent radiation-hydrodynamic simulations of larger clouds that produce hundreds of protostars have yielded populations of protostars whose mass distributions are statistically indistinguishable from the observed IMF \citep{bate12a, krumholz12a}.  Figure \ref{fig:bate12} shows some example results. This led a number of authors to conclude that gravity, hydrodynamics and thermal feedback may be the primary ingredients for producing the statistical properties of low-mass stars.  However, none of the above simulations included magnetic fields.

\bigskip
\noindent
\subsection{\textbf{Influence on the IMF: high-mass end}}
\bigskip
\label{ssec:thermal_highmass}

Thermal feedback from protostars is fundamentally a local effect, and is stronger for higher-mass protostars and/or greater accretion rates (equation \ref{eq:acclum}).  Therefore, it becomes even more important if the molecular cloud core has a high density and/or produces a massive protostar. In the absence of thermal feedback, a massive dense molecular cloud core is prone to fragment into many protostars since the Jeans mass scales inversely with the square root of the density.  Such fragmentation typically leads to a dense cluster of protostars that evolve according to competitive accretion, resulting in a cluster with a wide range of stellar masses \cite[e.g.][]{BonVinBat2004}.   However, the inclusion of thermal feedback can substantially alter this picture since the heating may be strong enough to exclude the vast majority of fragmentation, with the result that only a few massive stars are produced rather than a populous cluster \citep{krumholz07a}.  This implies that massive stars may be preferentially produced in regions with high densities \citep{KruMcK2008, Krumholzetal2010}. 

While thermal feedback can be important for reducing the level of fragmentation and producing massive stars, in some calculations it can be too dominant.  \cite{krumholz11c} found that as the star formation proceeds in a dense cluster-forming cloud and the thermal feedback becomes more intense, the rate of production of new protostars can decrease.  Since the protostars in the cloud continue to accrete more and more mass, this can lead to a situation in which the characteristic mass of the stellar population increases with time.  This means that the stellar mass distribution evolves with time, rather than always being consistent the observed IMF as in calculations of stellar clusters forming in lower-density clouds \citep{bate12a}.  In the long-term, this would result in a top-heavy IMF.  Combining the effects of radiative feedback and protostellar outflows may provide a way to reduce this effect, however (see Section \ref{sec:interactions}).

\citet{2010ApJ...711.1017P, 2010ApJ...719..831P, Petersetal2011} obtained a somewhat different result in their simulations, finding that radiative feedback only modestly reduced the degree of fragmentation. They term the phenomenon they observe fragmentation-induced starvation, a process by which, if a secondary protostar manages to form in orbit around a massive protostar it may reduce the growth rate of the massive protostar by accreting material that it would otherwise accrete. \citet{Girichidisetal2012, Girichidis12b} show that fragmentation-induced starvation also occurs in more general geometries, albeit in simulations that do not include radiative feedback. The relatively modest effects of radiative feedback in the \citeauthor{Petersetal2011}~simulations stands in contrast to the much stronger effects identified by \citet{krumholz11c, krumholz12b} and \citet{bate12a}.

Some of this difference may originate in the numerical method for treating radiation, with \citeauthor{Petersetal2011}~using a ray-tracing method that only follows photons directly emitted by the star, while \citeauthor{krumholz11c}~and \citeauthor{bate12a} use a diffusion method that follows the dust-reprocessed radiation but only indirectly treats direct stellar photons. However, a more likely explanation is a difference of initial conditions.  \citet{KruMcK2008} and \citet{Krumholzetal2010} argue that the surface density of a region is the key parameter that determines how effective radiative feedback will be, since it determines how effectively stellar radiation is trapped. In their simulations, \citet{krumholz11c, krumholz12b} consider a region similar to the center of the Orion Nebula Cluster, with a surface density $\Sigma \approx 1$ g cm$^{-2}$ ($\approx 5000$ $M_\odot$ pc$^{-2}$), while \citet{bate12a} simulates a region with $\Sigma \approx 0.2$ g cm$^{-2}$ ($\approx 1000$ $M_\odot$ pc$^{-2}$), similar to nearby low-mass star-forming regions such as Serpens or $\rho$ Ophiuchus. In contrast, \citet{2010ApJ...711.1017P, 2010ApJ...719..831P, Petersetal2011} use an initial condition with $\Sigma \approx 0.03$ g cm$^{-2}$ ($\approx 100$ $M_\odot$ pc$^{-2}$), comparable to the surface density in giant molecular clouds averaged over $>10$ pc scales (see the chapter by \textit{Dobbs et al.}~in this volume). \citeauthor{Petersetal2011}'s surface density is low enough that their cloud is optically thin in the near-infrared, and it is not surprising that radiative heating has minimal effects in such an environment, since any stellar radiation absorbed by the dust escapes as soon as it is re-emitted, rather than having to diffuse outward and heat the cloud in the process.

Simultaneous observations of fragmentation and temperature distribution in cluster forming clumps can provide a key observational test of the numerical results. Recent high angular resolution observations of dense, massive infrared dark clouds, precursors to cluster forming regions, have begun to detect massive cores at the early phases of cluster formation \citep{rathborne2008, zhang2009, bontemps2010, zhang2011, wang2011, Longmoreetal2011, wang2012}. Indeed, these cores contain masses at least a factor of 10 larger than the Jean mass \citep{zhang2009}. However, they do not appear to 
lie in  the high temperature sections of the clump \citep{wang2012}, which appears to conflict with numerical simulations. Future observations of larger sample of massive cluster forming clumps will provide a statistical trend that further constrains the role of thermal feedback to the formation of massive stars (see Section \ref{ssec:futureobs}).

\bigskip
\section{\textbf{PUTTING IT ALL TOGETHER}}

\bigskip
\subsection{\textbf{Interactions Between Feedback Mechanisms}}
\label{sec:interactions}
\bigskip

\subsubsection{Combined effects of multiple mechanisms}

In the preceding sections, we have discussed the effects of various different types of feedback individually.  However, in reality, feedback mechanisms frequently act simultaneously.  For low-mass star formation, the dominant mechanisms are thought to be protostellar outflows and thermal feedback.  Radiation pressure is negligible and supernovae do not occur. Photoionization will only affect the immediate vicinities of protostars, though it may be crucial for the erosion of protoplanetary discs \citep{Hollenbachetal1994, YorWel1996, RicYor1997, ClaGenSot2001}.  By contrast, the feedback from high-mass protostars involves all of the above mechanisms.  

Both analytic and numerical investigations of multiple mechanisms are few. \citet{mckee84a} considered the interaction of a stellar wind and photoionization with a clumpy medium, and concluded that the photoevaporation of clumps would control the dynamics, either by creating an ionized medium that would pressure-confine the wind or by providing a mass load that would limit its expansion. \citet{krumholz05_outflows} showed that  protostellar outflows would significantly weaken the effects of radiation pressure feedback by creating escape routes for photons.  \citet{Cunninghametal2011} confirmed this prediction with simulations, and also showed that focusing of the radiation by the outflow cavity prevents the formation of radiation-pressure driven bubbles and the associated Rayleigh-Taylor instabilities seen in earlier calculations \citep{krumholz09c}.

\citet{HanKleMcKFis2012} investigated the combined effects of radiative transfer and protostellar outflows on low mass star formation.  As the outflows reduce the accretion rates of the protostars, they also reduce their masses and luminosities and hence the level of thermal feedback.  They found that the outflows did not have a significant impact on the kinematics of the star forming cloud, but the calculations did not include magnetic fields.  In Section \ref{ssec:thermal_highmass}, we mentioned that \cite{krumholz11c} found that in massive dense star-forming clouds, thermal feedback could be so effective at inhibiting fragmentation that it led to a top-heavy IMF.  However, \citet{krumholz12a} showed that this ``over-heating problem" could be reduced by including the effects of large-scale turbulent driving and protostellar outflows, since both of these processes lower protostellar accretion rates and thus the effects of thermal feedback.  This is one case where the details of, and uncertainties in, accurately determining the luminosities of protostars (see Section \ref{ssec:feedbackorigins}) can play a crucial role in the outcome of star formation.

\red{A number of authors have also simulated the interaction of stellar winds with photoionized regions \citep[e.g.][]{garcia-segura96b, garcia-segura96c, freyer03a, freyer06a, toala11a}. However, these simulations generally begin with a single star placed in a uniform, non-self-gravitating medium. While this setup is useful for studying the internal dynamics of wind bubbles, it limits the conclusions  that can be drawn about how the feedback affects the formation of star clusters, where the surrounding medium is highly structured and strongly affected by gravity.}

\subsubsection{Feedback, turbulence, and magnetic fields}

Some feedback effects are enhanced by the presence of turbulence and/or magnetic fields, while others are reduced.  For example, \citet{banerjee07a} examined jets propagating into a quiescent medium and concluded that they do not drive supersonic turbulence, while \citet{cunningham09a} shows that jets propagating into a pre-existing turbulent medium could inject energy into the turbulence, thus potentially allowing outflows to sustain existing turbulence in a star-forming region.  It has been noted, however, that the velocity power spectrum of the turbulence generated by outflows in magnetized clouds may be slightly steeper than that generated by isotropic forcing \citep{carroll09a}.

By themselves, stronger magnetic fields have been shown to reduce the star formation rate in molecular clouds \citep{nakamura07a, PriBat2008, PriBat2009}. However, magnetic fields can also interact with feedback effects in subtle ways. We already discussed (Section \ref{sssec:outflowtheory}) how magnetic fields enhance the effects of protostellar outflows by raising the efficiency with which they deposit their energy in clouds. \citet{2012ApJ...745..158G} demonstrated a similar magnetic enhancement in the effects of photoionization feedback. On the other hand, \citet{Petersetal2011} investigate the interaction of magnetic fields with ionizing radiation on smaller scales, and find that ionizing radiation tends to make it harder for massive stars to drive collimated magnetized outflows, because the pressure of the photoionized gas disrupts the magnetic tower configuration that can drive outflows from lower mass stars.

In a similar vein, while it has been known for some time that magnetic fields alone reduce fragmentation in collapsing gas \citep[e.g.][]{Hennebelleetal2011}, they also appear to greatly enhance the effects of thermal feedback. \citet{Commerconetal2010, Commerconetal2011} and \citet{myers13a} find that magnetic fields provide an efficient mechanism for angular momentum transport and thus tend to increase accretion rates onto forming stars. This in turn raises their accretion luminosities, and thus the strength of thermal feedback. Moreover, thermal feedback and magnetic fields work well in combination to reduce fragmentation in both low-mass and high-mass star formation calculations. For low-mass clusters, \cite{PriBat2009} found that thermal feedback inhibits fragmentation on small scales, while magnetic fields provide extra support on large-scales. The result is that the combination of magnetic fields and thermal feedback is much more effective than one would naively guess. For massive cores, \citet{Commerconetal2010, Commerconetal2011} find a similar effect operating at early times, up to the formation of a Larson's first core. However, they are unable to address the question of fragmentation at later times.

\begin{figure*}
\epsscale{1.5}
\plotone{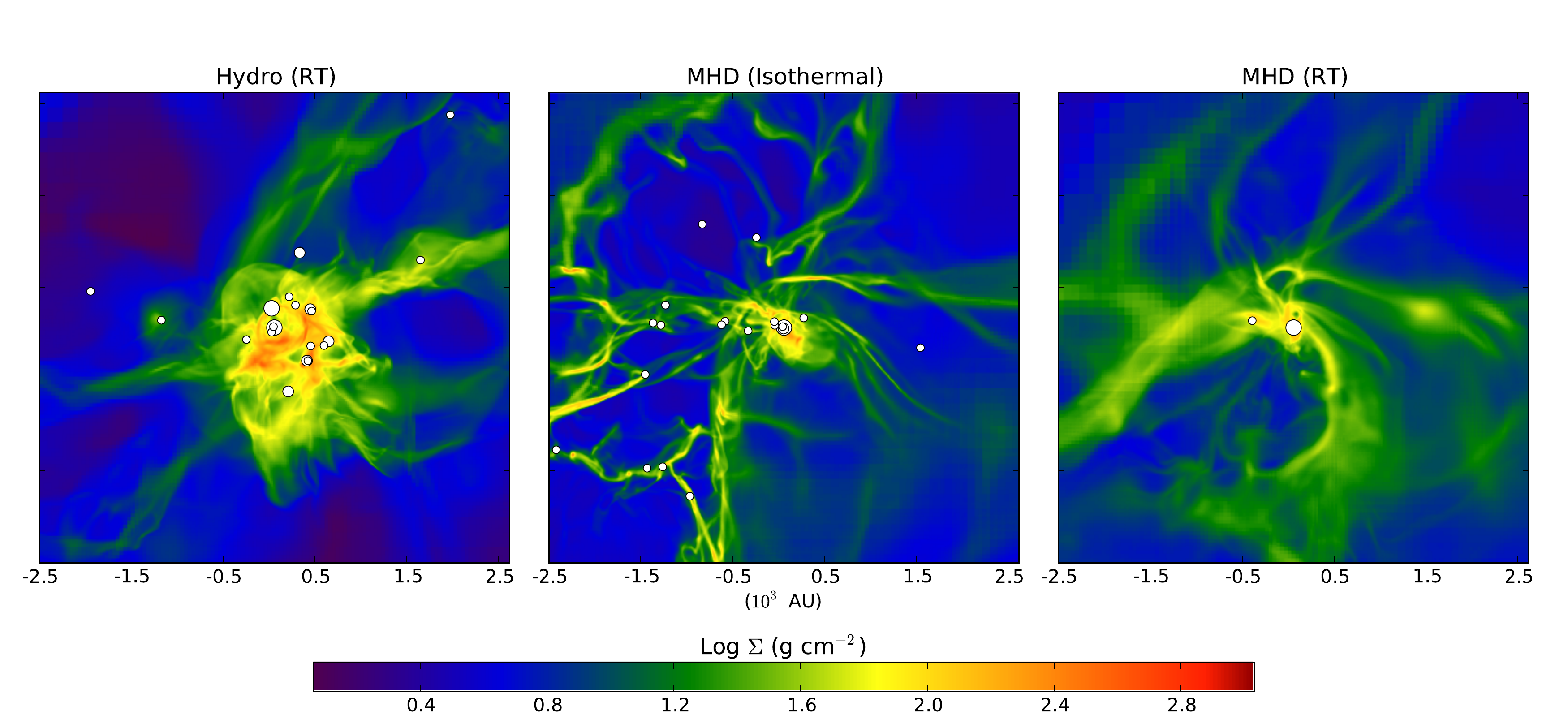}
\epsscale{1.0}
\caption{\small
\label{fig:myers13}
Results from three simulations by \citet{myers13a}. The color scale shows the column density, and white circles show stars, with the size of the circle indicating mass: $<1$ $M_\odot$ (small cirlces), $1-8$ $M_\odot$ (medium circles), and $>8$ $M_\odot$ (large circles). The panel on the left shows a simulation including radiative transfer (RT) but no magnetic fields, the middle panel shows a simulation with magnetic fields but no radiative transfer, and the rightmost panel shows a simulation with both radiative transfer and magnetic fields. All runs began from identical initial conditions, and have been run to 60\% of the free-fall time at the initial mean density.}
\end{figure*}

\citet{myers13a} use a subgrid stellar evolution model that allows them to run for much longer times than \citeauthor{Commerconetal2010}, and find that thermal feedback inhibits fragmentation in the dense central regions, while magnetic fields inhibit it in the diffuse outer regions.  They find that strong magnetic fields and thermal feedback in massive dense cores make it very difficult to form anything other than a single massive star or, perhaps, a binary. Figure \ref{fig:myers13} illustrates this effect. It shows the results of three simulations by \citet{myers13a} using identical resolution and initial conditions, one with magnetic fields but no radiation, one with radiation but no magnetic fields, and one with both radiation and magnetic fields. The run with both forms many fewer stars than one might naively have guessed based on the results with radiation or magnetic fields alone.

These results are in contrast to those of \citet{Petersetal2011}, who also include both radiation and magnetic fields, but find only a modest reduction in fragmentation. As discussed in Section \ref{ssec:thermal_highmass}, this is likely a matter of initial conditions: \citeauthor{Commerconetal2010}~and \citeauthor{myers13a}~consider dense prestellar cores with surface density $\Sigma \approx 1$ g cm$^{-2}$ (chosen to match observed infrared dark cloud cores -- \citealt{swift09a}), while \citet{Petersetal2011} simulate much more diffuse regions with $\Sigma \approx 0.03$ g cm$^{-2}$, and is not clear if their calculations ever evolve to produce the sorts of structures from which \citeauthor{Commerconetal2010}~and \citeauthor{myers13a}~begin.

The combined effects of magnetic fields and thermal feedback can even modify star formation rates. Thermal feedback by itself reduces star formation rates by at most tens of percent \citep{bate09a, bate12a, Krumholzetal2010}, but adding magnetic fields can reduce the star formation rate in low mass clusters by almost an order of magnitude over purely hydrodynamic collapse \citep{PriBat2009}, and significantly more than magnetic fields alone.

\bigskip
\subsection{\textbf{Triggering}}
\label{ssec:triggering}
\bigskip

Thus far we have primarily focused on negative feedback, in the sense of restraining or terminating star formation. However, it is also possible for feedback to be positive, in the sense of promoting or accelerating feedback. The statistical arguments outlined in Section \ref{ssec:why} would tend to suggest that negative feedback must predominate, but this does not necessarily imply that positive feedback never occurs or cannot be important in some circumstances.


Positive feedback is usually referred to as triggered or induced star formation. This phrase can mean increasing the star formation rate, increasing the star formation efficiency, or increasing the total number of stars formed. These definitions can all be applied locally or globally. \citet{2007MNRAS.377..535D} draw a distinction between weak triggering -- temporarily increasing the star formation rate by inducing stars to form earlier -- and strong triggering -- increasing the star formation efficiency by causing the birth of stars that would not otherwise form. They note that it may be very difficult for observations to distinguish these possibilities.

Analytic studies by \citet{1994A&A...290..421W} suggest that the gravitational instability operating in swept-up shells driven into uniform gas by expanding H~\textsc{ii} regions or wind bubbles should be an efficient triggering process. \citet{1994MNRAS.268..291W} extended this work to show that this process should result in a top-heavy IMF, a result also found in simulations of fragmenting shells by \red{\citet{wunsch10a} and} \citet{2011MNRAS.411.2230D}. This raises the intriguing prospect of star formation as a self-propagating process \citep[e.g.][]{1981ApJ...249...93S,1983ApJ...265..202S}. However, simulations of ionizing feedback in fractal \red{\citep{walch13a}} and turbulent clouds \citep{2007MNRAS.377..535D,2012MNRAS.422.1352D,2012MNRAS.427.2852D,2013MNRAS.431.1062D} suggest that this is not the case. They find that, while ionization feedback can modestly change the rate, efficiency and number of stars, it does not significantly alter the IMF.

Pillars or ``Elephant Trunks" are a widespread and distinctive feature of star-forming regions and have often been invoked as signposts of triggering \citep[e.g.,][]{2000ApJ...532L.145S,2005AJ....129..888S,2010ApJ...712..797B}. \citet{2001MNRAS.327..788W}, \citet{2006MNRAS.369..143M}, \red{\citet{gritschneder10a}}, \citet{2011MNRAS.412.2079M}, \citet{2012A&A...538A..31T,2012A&A...546A..33T}\red{, and \citet{walch13a}} have simulated pillar formation in a wide variety of initial conditions and provide several plausible mechanisms for their origins. However, it is not clear to what extent these morphological features are actually indicative of triggered star formation.

There is also large body of literature on the induced collapse of initially-stable density configurations, such as Bonner-Ebert spheres, by winds or by H~\textsc{ii} regions, known as radiation-driven implosion  \citep[e.g.][]{1982ApJ...260..183S,1984ApJ...282..178S,1983ApJ...271L..69K,bertoldi89a, bertoldi90a,2003MNRAS.338..545K,2011ApJ...736..142B}. This process is able to produce not only single stars, but small groups, and since the initial conditions are stable by construction, this is a good example of strong triggering.

Establishing the occurrence of triggered star formation analytically or numerically is relatively straightforward. One can use either initial conditions that are stable in the absence of feedback, as in the radiation-driven implosion simulations, or, for more complex initial conditions, control simulations without feedback. Detecting triggering in observed systems is much more difficult, since neither of these paths are open to the observer. Instead, observers must search for circumstantial evidence for triggering. Examples of such attempts in the literature include surveys of young stars near bubbles, ionization fronts or bright--rimmed clouds \citep[e.g.][]{2007A&A...467.1125U,2008A&A...482..585D,2009ApJ...700..506S,2010MNRAS.406..952S} or pillars, campaigns to find clusters elongated towards feedback sources or showing strong age gradients in young stars \citep[e.g.][]{1995ApJ...455L..39S,2009MNRAS.396..964C,2012MNRAS.426.2917G}, and searches for regions of unusually high star formation rate or efficiency \citep[e.g.][]{2011ApJ...736..142B}. In these observational campaigns, the standard practice has been to claim evidence for triggered star formation whenever there are coherent structures within which the stellar age differences are smaller than the crossing time of the cloud structure. Generally such structures are compressed shells of molecular gas with tens of pc radii, such as the CepOB2 bubble \citep{1998ApJ...507..241P} or the MonR2 GMC \citep{1994ApJ...430..252X}.

\red{This approach must be treated with caution, because star formation correlates with the presence of dense gas on pc scales even in the absence of obvious feedback or triggering \citep{2010ApJ...723.1019H,2011ApJ...739...84G}. Thus one expects to find correlations between YSOs and nebulosity features even in the absence of triggering. For evidence of triggering to be convincing, one must show that the correlation between star formation and feedback-driven features is in excess of what one would expect simply from the baseline correlation of YSOs and dense gas. For example, in some cases the ratio of YSO mass to molecular gas mass is far higher than one normally finds in active star-forming regions \citep[e.g.][]{2009ApJ...699.1454G}. However, even in such cases it is unclear whether the enhancement is a matter of positive feedback (enhanced star formation raising the stellar mass) or negative feedback (ablation of gas reducing the gas mass without creating any additional stars).}

The contribution of triggered star formation to the global star formation rate is unclear. Most GMCs have uniformly low star formation efficiencies \citep{evans09a}, showing no obvious correlation with numbers of OB stars or signs of feedback such as bubbles. \cite{2012ApJ...755...71K} and \cite{2012MNRAS.421..408T} use statistical correlations between large catalogues of bubbles and young stellar objects to infer that triggered star formation may contribute at most tens of percent to the Galactic population of massive stars. Such a small variation is well within the cloud-to-cloud scatter in star formation efficiency seen by \citet{evans09a}; indeed, it is smaller than the typical observational uncertainty on the star formation rate, particularly within a single cloud. Similarly, simulations by \citet{2012MNRAS.427.2852D,2013MNRAS.431.1062D} imply that, while triggering certainly does occur, it is overwhelmed by negative feedback on the scale of GMCs.

Ultimately, quantifying the sizes and timescales for the destructive influences of feedback is essential; outside of these zones of negative influence, large scale effects from feedback may indeed turn positive, especially in aggregate with other moderately nearby feedback sources.  Are these aggregate effects truly triggering, or are they simply the source of the large-scale turbulence that drives the creation of the next generation of star-forming molecular material?

\bigskip
\section{\textbf{FUTURE PROSPECTS}}
\bigskip

\subsection{\textbf{Observations}}
\bigskip
\label{ssec:futureobs}

In order to place more stringent constraints on how feedback
affects the star and cluster formation process, new observations must tackle
the problem on two fronts.  Naturally, we must continue to push to observe
statistically significant samples of  ever more extreme (and more distant)
environments as they come within reach of new facilities
across the EM spectrum.  In addition, we must endeavor to improve our
understanding of star formation within nearby, low-feedback environments in
order to set an interpretive framework that will be essential for
determining the net effect of more extreme feedback.  In both cases,
characterization of the properties and kinematics of star-forming gas and the
forming stars themselves will be essential.

High angular resolution observations of protocluster-forming regions have
revealed fragmentation of dense molecular gas at spatial scales of $ < 10^3$ AU
\citep{rathborne2008,zhang2009,bontemps2010,zhang2011,wang2011,Longmoreetal2011,Wangetal2012,palau2013}.
Despite limited sensitivity and dynamic range, these observations provided
valuable snapshots of density and temperature distributions of the star forming
gas.  The next generation of
interferometers such as ALMA will provide  even more detailed images of density
distribution and linewidth of star forming cores in protocluster forming
molecular clumps. At the same time, the temperature distribution within the
molecular clump can be readily obtained from observations with Karl Jansky VLA
\citep[e.g.][]{wang2008,Wangetal2012}. The combined information will constrain the
thermal dynamic properties and fragmentation of cluster forming clumps,
providing direct information on the initial physical properties of molecular
gas that gives rise to a stellar cluster.

A survey of nearby clusters,  within $\sim 500$ pc, using existing millimeter
interferometers \citep[e.g., CARMA, see][]{plunkett13a} or heterodyne receiver
arrays in (sub)-mm (single-dish) telescopes  (e.g., SEQUOIA on LMT or the
64-pixel array Supercam on SMT) allow for efficient mapping of outflows using
different CO transition lines. Even though in many cases the angular resolution
will not be enough to resolve individual outflows in regions with a high
density of protostars, these observations will be necessary to obtain total
molecular outflow energetics and compare them with cloud energy and gas
distribution for a large sample of clusters. Observations of these regions
should also include maps of higher (column) density tracers, such as $^{13}$CO
and C$^{18}$O in order to investigate the impact of outflows and winds in the
cloud structure and turbulence. 

In the near future, when ALMA is completed and on-the-fly (OTF) mapping
becomes available, it will be feasible to conduct studies similar to the ones
described above for a (larger) sample of clusters (within about 10 kpc) that is
representative of the galactic cluster population. High-resolution ALMA
observations of massive (far-away) clusters will also help in studying the
impact of compact H II regions on the surrounding molecular cloud. In addition,
multi-epoch VLA continuum observations of compact H II regions will allow
measurement of the expansion velocity of the ionized bubble, and in concert
with multi-wavelength data, will place constrain on the impact of winds and
radiation pressure on the surrounding cluster environment. 
However, these observations may be complicated by the fact that, at
least during the hypercompact stage, some HII regions are observed to
shrink rather than expand \citep{galvan-madrid08a}, likely as a result of
the motions of dense material near the ionizing source causing part of
the H II region to be shadowed \citep{galvan-madrid11a}.

At larger scales, OTF observations using heterodyne receiver arrays in
millimeter single-dish telescopes will allow fast mapping of molecular gas that
has been swept-up by bubbles and SNR in high-mass star forming regions. In
lower density regions, where there is little or no molecular gas, large-scale
galactic HI surveys with SKA precursor telescopes, like GASKAP
\citep{dickey13a} will provide useful information on the effects of stellar
feedback on the low-density outskirts of clusters.

Confidently obtaining the census of low mass young stars forming in more
diverse feedback-affected environments is essential to establishing correlations
between feedback sources (e.g.~their nature, position, and intensity) and
changes in star-gas column density correlation or other aspects of the star formation process.  Given the simultaneous
needs of moderate to high extinction penetration and membership isolation from
field stars, infrared and X-ray imaging capabilities remain the central means
for identifying YSOs.  Recent infrared YSO membership surveys with {\it Spitzer} have
considerably expanded our knowledge of YSOs in the nearest kpc.
However, they are limited in their ability to discern members projected on
bright nebulosity, thus only a relatively narrow range of galactic environments
have been thoroughly searched for forming stellar content throughout the
stellar mass range \citep{allen07,evans09a,gutermuth09a,megeath2012}.   

X-ray observations, particularly deep and high resolution imaging with Chandra,
have proven effective at bypassing the nebulosity limitations of IR
surveys, detecting substantial numbers of YSOs in particularly IR-bright high
mass star forming regions.  X-ray emission from YSOs is generally considered to
be a product of magnetic field activity, and thus not strongly correlated with
the presence of a disk \citep{feigelson07}.  The resulting YSO census derived
from X-ray imaging therefore trades the disk bias of IR surveys for a broad
completeness decay as a function of luminosity, as stochastic flaring events of
gradually increasing strength, and therefore rarity, are required to detect
lower mass sources with smaller quiescent luminosities.  

In the near term, improving capabilities in adaptive optics on ever larger
aperture ground-based optical and near-IR telescopes will facilitate other
means of young stellar membership isolation, both via facilitating high
throughput spectroscopy as well as yielding sufficient astrometric precision
for proper motion characterization \citep[e.g.][]{lu13}.  Looking further
ahead, the considerable improvement in angular resolution and sensitivity at
near- and mid--IR wavelengths afforded by JWST will dramatically improve the
contrast between nebulosity and point sources in Milky Way star forming
regions, enabling Spitzer-like mid-IR YSO surveys out to much greater distances
and in regions influenced by much more significant feedback sources.  The
resulting censuses of young stars with IR-excess will reach well down the
stellar mass function in regions found both within the Molecular Ring as well
as toward the outskirts of the Milky Way.  Unfortunately, next generation X-ray
space telescopes of similar angular resolution and better sensitivity relative
to Chandra (e.g. AEGIS, AXSIO, SMART-X) remain in the planning phase for launch
on a time frame outside the scope of this review.

\bigskip
\noindent
\subsection{\textbf{Simulations and Theory}}
\bigskip

In the area of simulations and theory, in the next few years we can expect improvements in several areas. The first need that should be apparent from the preceding discussion is for simulations that include a number of different feedback mechanisms, and that can assess their relative importance. As of now, there are simulations and analytic models addressing almost every potentially-important form of feedback: pre-main sequence outflows, main sequence winds, ionizing radiation, non-ionizing radiation, radiation pressure, and supernovae. However, no simulation or model includes all of them, and few that include more than one. Moreover, many simulations including feedback do not include magnetic fields. As discussed in Section \ref{sec:interactions}, interactions between different feedback mechanisms, and between feedback and pre-existing turbulence and magnetic fields, are potentially important, but remain largely unexplored.

This limitation is mostly one of code development. Designing and implementing the numerical methods required to treat even one form of feedback in the context of an adaptive mesh refinement, smoothed particle hydrodynamics, or other code capable of the high dynamic range required to study star formation is the work of multiple years. As a result, no one code includes treatments of all the potentially important mechanisms. However, that situation is improving as code development progresses, and the pace of development is increasing as at least some of the remaining work involves porting existing techniques from one code to another, rather than developing entirely new ones. By the time of the next Protostars and Planets review, it seems likely that there will be more than a few published simulations that include MHD, protostellar outflows, main sequence winds, and multiple radiation effects, including pressure and heating by both ionizing and non-ionizing photons. There are also likely to be improvements in the numerical techniques used for many of these processes, particularly radiative transfer, where it seems likely that in the next few years many codes will be upgraded to use variable Eddington tensor methods \citep[e.g.][]{davis12a, jiang12a} or other high order methods such as $S_n$ transport.

A second major area in need of progress is the initial conditions used in simulations of star cluster formation. At present, most simulations begin with highly idealized initial conditions: either spherical regions that may or may not be centrally concentrated, or turbulent periodic boxes. In simulations without feedback, \citet{girichidis11a, Girichidisetal2012, Girichidis12b} show that the results can depend strongly on which of these setups is used. For simulations with feedback, those that run long enough and contain forms of feedback such that they are able to reach a statistical steady state are probably fairly insensitive to the initial conditions. For the vast majority of simulations, though, particularly those where gas expulsion is rapid and occurs at most $\sim 1$ dynamical time after the onset of star formation, the initial conditions likely matter a great deal. In reality, the dense regions that form clusters are embedded within larger giant molecular clouds, which are themselves embedded in a galactic disk. They likely begin forming stars while they are still accreting mass, and the larger environments that are missing in most simulations can provide substantial inputs of mass, kinetic energy, and confining pressure. Simulations of that include both the formation of a cloud and feedback have begun to appear in the context of studies of giant molecular clouds on larger scales \citep[e.g.][]{vazquez-semadeni10a}, and there is clearly a need to extend this approach to the smaller, denser scales required to study the formation of star clusters.

A third, closely related problem is that the current generation of simulations usually explore a very limited range of parameters -- for example, only a single cloud mass and density, or a single magnetic field strength or orientation. As a result, it is difficult to draw general conclusions, particularly when results differ between groups. For example, \citet{wang10a} find that including protostellar outflows dramatically reduces the star formation rate in their simulations of cluster formation, while \citet{krumholz12b}, using essentially the same prescription to model outflows, find that the effects on the star formation rate are much more modest. Is this because \citeauthor{wang10a}'s simulations include magnetic fields and those of \citeauthor{krumholz12b}~do not? Because \citeauthor{krumholz12b}'s simulations include radiation and \citeauthor{wang10a}'s do not? Or because \citeauthor{wang10a}~simulate a region modeled after a relatively low density region like NGC 2264, while \citeauthor{krumholz12b}~choose parameters appropriate for a much higher density region like the core of the Orion Nebula Cluster? Since each paper simulated only a single environment, the answers remain unknown. While there are analytic models that provide some guidance as to which feedback mechanisms might be important under what conditions \citep[e.g.][]{fall10a}, there are precious few parameter studies. (See \citet{Krumholzetal2010}, \citet{myers11a} and \citet{2013MNRAS.430..234D} for some of the few exceptions.) This will need to change in the coming years.
 
All of these advances are likely to require fundamental changes in the algorithms and code architecture used for numerical simulations of star cluster formation. Using present algorithms, parallel simulations of star cluster formation that go to resolutions high enough to (for example) resolve fragmentation to the IMF, and that include even one or two feedback mechanisms, often require many months of run time. Adding more physical processes, or more accurate treatments of the ones already included, will only exacerbate the problem. Part of the problem is that the techniques currently in use do not scale particularly well on modern massively-parallel architectures. This is partly a matter of physics: the problem of star formation is inherently computationally difficult due to the wide range of time and spatial scales that must be treated. However, it is also partly a matter of code design: few modern multi-physics codes have been ported to hybrid threaded / message-passing architectures, and even fewer have been optimized to run on GPUs or similar special-purpose hardware. In addition to a improving the physics in our codes, a great deal of software engineering will be required to meet the goals laid out above in time for the next Protostars and Planets review.

\textbf{ Acknowledgments.} 
This work was supported by the following sources:
the Alfred P.~Sloan Foundation (MRK); 
National Science Foundation grants AST-0955300 (MRK),
AST-0845619 (HGA), AST-1313083 (ZYL), AST-0908553 (RIK), and NSF12-11729 (RIK);
NASA ATP grants NNX13AB84G (MRK and RIK), 
and NNX10AH30G (ZYL);
NASA ADAP grants NNX11AD14G (RAG) and NNX13AF08G (RAG);
NASA through Chandra Award Number GO2-13162A (MRK) issued by the Chandra X-ray Observatory Center, which is operated by the Smithsonian Astrophysical Observatory for and on behalf of the National Aeronautics Space Administration under contract NAS8-03060;
NASA through Hubble Award Number 13256 (MRK) issued by the Space Telescope Science Institute, which is operated by the Association of Universities for Research in Astronomy, Inc., under NASA contract NAS 5-26555;
NASA by Caltech/JPL awards 1373081 (RAG), 1424329 (RAG), and 1440160 (RAG) in support of {\it Spitzer Space Telescope} observing programs;
the US Department of Energy at the Lawrence
Livermore National Laboratory under contract DE-AC52-07NA2734 (RIK);
the DFG cluster of excellence ``Origin and Structure of the Universe" (JED);
the hospitality of the Aspen Center for Physics, which is supported by the National Science Foundation Grant PHY-1066293 (MRK, RIK, and RAG).

\bigskip

\bibliographystyle{ppvi_lim1.bst}
\bibliography{refs}

\end{document}